\documentclass[aps,prd,preprintnumbers,amsmath,amssymb,latexsym,nofootinbib,array,enumerate,letter,superscriptaddress, floatfix]{revtex4-2}
\usepackage{blindtext}
\usepackage{lineno}
\usepackage{bm,color,xcolor}
\usepackage{slashed} 
\usepackage{graphicx}
\usepackage{subfigure} 
\usepackage{amsmath}
\usepackage{amssymb}
\usepackage{multirow}
\usepackage{float}
\usepackage{comment}
\usepackage{booktabs}
\usepackage{rotating}
\usepackage{fontawesome} \usepackage{mathtools}
\usepackage[colorlinks=true
,urlcolor=blue
,anchorcolor=blue
,citecolor=blue 
,filecolor=blue
,linkcolor=blue
,menucolor=blue
,linktocpage=true
,pdfproducer=medialab
,pdfa=true
]{hyperref}
\usepackage{lipsum}

\makeatletter
\def\l@subsubsection#1#2{}
\makeatother

\newcommand{\lnu}{\lambda_{\phi\nu}}

\newcommand{\CnuB}{\mathrm{C}\nu\mathrm{B}}

\begin{document}

\title{SN1987A constraints on the neutrino-dark-fermion interaction from resonant scattering with $\CnuB$}

\author{Christina Gao} 
\email{gaoy3@sustech.edu.cn}
\affiliation{Department of Physics, Southern University of Science and Technology, Shenzhen, 518055, China}

\author{Ke Hu} 
\email{12531056@mail.sustech.edu.cn}
\affiliation{Department of Physics, Southern University of Science and Technology, Shenzhen, 518055, China}

\author{Kun-Feng Lyu} 
\email{kunfeng.lyu@ou.edu}
\affiliation{Homer L. Dodge Department of Physics and Astronomy, University of Oklahoma, Norman, OK 73019, USA}

\author{Vincent Zou} 
\email{zour2025@mail.sustech.edu.cn}
\affiliation{Department of Physics, Southern University of Science and Technology, Shenzhen, 518055, China}

\begin{abstract}
Neutrino self-interactions mediated by a light scalar offer a compelling resolution to cosmological tensions and may naturally arise in neutrino mass generation mechanisms. When the scalar also couples to a light dark-sector fermion, supernova neutrinos can resonantly annihilate with the cosmic neutrino background (C$\nu$B) into invisible dark radiation, depleting the flux en route to Earth. We use this depletion to constrain the neutrino-scalar coupling from the SN1987A data. Within a Bayesian framework we analyze the data for two supernova neutrino emission models, a parameterized model and a 2D hydrodynamic simulation, and for two coupling types, a mass-independent one and a mass-proportional one. 
Our results show that new-physics limits from SN1987A cannot be quoted independently of the heavy-flavor emission, the flavor conversion, or the coupling structure. Finally, we forecast that a future high-statistics burst recorded by Hyper-Kamiokande would restore a meaningful upper bound even when flavor conversion is included.
\end{abstract}

\maketitle

\setcounter{secnumdepth}{3}
\setcounter{tocdepth}{1}

%{\bf Introduction}---%
%
\section{Introduction}

Neutrinos, the lightest fermions in the Standard Model (SM), remain one of the least understood sectors of particle physics. Their feeble interactions leave ample room for couplings beyond the SM. Neutrino self-interactions ($\nu$SI) mediated by a light scalar $\phi$, for instance the Majoron, the pseudo-Nambu--Goldstone boson of spontaneously broken lepton number~\cite{Gelmini:1980re,Chikashige:1980ui,Aulakh:1982yn,Joshipura:1992hp}, have been widely studied~\cite{Berryman:2022hds}.
Beyond their connection to the origin of neutrino masses and to dark-sector physics, such interactions have been invoked to ease the Hubble and $\sigma_8$ tensions~\cite{Kreisch:2019yzn,Ghosh:2019tab,Das:2020xke,RoyChoudhury:2020dmd,Brinckmann:2020bcn,Venzor:2023aka} and leave characteristic imprints in the matter power spectrum~\cite{Kumar:2022vee,Libanore:2025ack,Camarena:2023cku,He:2023oke,Pal:2024yom,Poudou:2025qcx,He:2025jwp,Noriega:2025ulc}. 

Neutrino self-interactions are, however, tightly constrained by terrestrial experiments and astrophysical probes. Rare meson decays and neutrinoless double-beta-decay searches~\cite{Lessa:2007up,Agostini:2015nwa,Pasquini:2015fjv,Blum:2018ljv,Berryman:2018ogk,Brune:2018sab, NA62:2021bji,deGouvea:2019qaz,Brdar:2020nbj, PIENU:2021clt,Kharusi:2021jez,Dev:2024ygx,Zhang:2024meg}, supernova observations~\cite{Manohar:1987ec,Farzan:2002wx, Heurtier:2016otg,Das:2017iuj,Shalgar:2019rqe, Akita:2022etk, Chang:2022aas, Fiorillo:2022cdq, Akita:2023iwq}, high-energy astrophysical neutrinos~\cite{Ioka:2014kca,Ng:2014pca,Bustamante:2020mep,Esteban:2021tub}, and Big Bang nucleosynthesis (BBN)~\cite{Huang:2017egl, Huang:2021dba} together exclude flavor-universal couplings and leave only narrow flavor-dependent windows~\cite{Blinov:2019gcj,Lyu:2020lps}. Recently, Ref.~\cite{Das:2025asx} proposed to couple the scalar mediator $\phi$ to an additional massless dark sector fermion $\chi$ with an $\mathcal{O}(1)$ coupling strength. Active neutrinos can then resonantly convert into dark radiation between BBN and recombination. Because the self-interacting dark radiation behaves like neutrinos at the CMB epoch, the cosmological signatures of $\nu$SI are reproduced while the required neutrino-scalar coupling $\lambda_{\phi\nu}$ can be far smaller than in the conventional $\nu$SI scenario, relaxing the laboratory and astrophysical bounds. 

\begin{figure}
    \centering
    \includegraphics[width=0.6\linewidth, trim= 100 50 0 100]{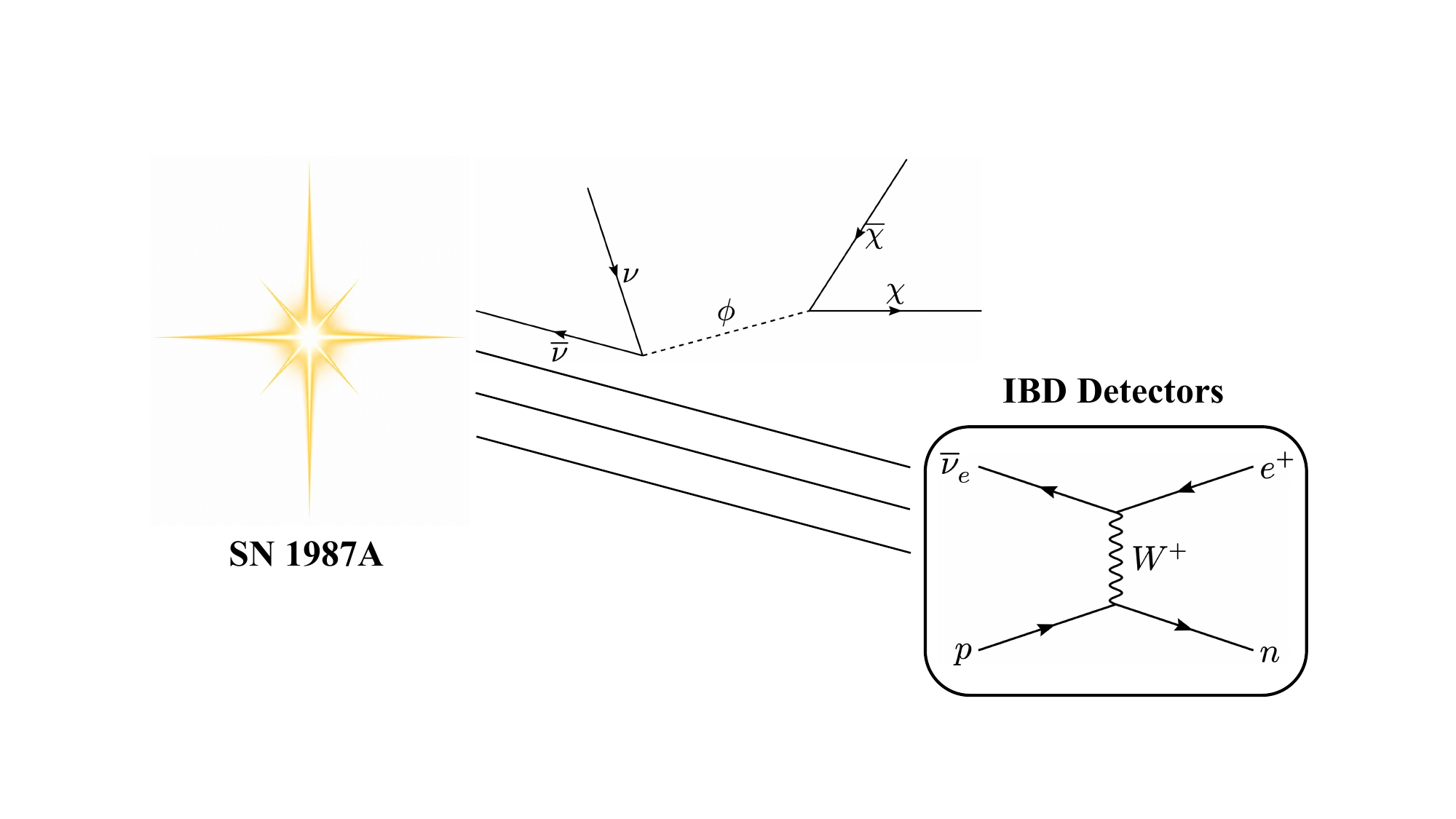}
    \caption{The scattering between SN neutrinos and cosmic neutrino background can resonantly convert to the scalar mediator, subsequently decaying to dark radiation. This phenomenon could induce an attenuation of SN neutrino flux, which can be detected by ground-based detectors with IBD process.}
    \label{fig:Schematic_Diagram}
\end{figure}

This raises the question of whether such a feeble $\lambda_{\phi\nu}$ coupling can still be probed. We show that it can, using the relic neutrinos that pervade the Universe. The cosmic neutrino background (C$\nu$B) is a firm prediction of standard cosmology~\cite{Lesgourgues:2006nd} but has so far eluded direct detection~\cite{Weinberg:1962zza,Long:2014zva,PTOLEMY:2019hkd}. One of the few ways it may become observable is by scattering energetic neutrinos from astrophysical sources~\cite{He:2025bex,Esteban:2021tub,Bharadwaj:2026ufv,Brdar:2022kpu,Machado:2025ltu,Wang:2025qap}. In this work, we consider a scenario where a supernova (SN) antineutrino annihilates with a relic neutrino through an $s$-channel scalar into a pair of dark fermions, $\nu\bar\nu\to\phi\to\chi\bar\chi$. When the center-of-mass energy sits on the resonance, this process efficiently depletes the SN neutrino flux en route to Earth (Fig.~\ref{fig:Schematic_Diagram}). The mechanism is closely analogous to the Glashow resonance~\cite{Glashow:1960zz,Barger:2014iua,Huang:2019hgs,IceCube:2021rpz}, the resonant formation of a $W$ boson when an ultra-energetic antineutrino strikes a non-relativistic electron. The historic neutrino burst from SN1987A remains the only direct observation of core-collapse SN neutrinos, and it is the target of our analysis. Whereas previous SN limits on $\lambda_{\phi\nu}$ are cooling bounds, set by the energy that bremsstrahlung-produced scalars carry out of the core, we instead constrain the coupling through the attenuation the flux suffers during propagation.

Extracting this bound requires a model of the SN neutrino emission. Two complementary approaches have been developed in the literature: parameterized analytic models~\cite{Loredo:2001rx, keil2003monte, pagliaroli2009improved, Tamborra:2012ac} and multidimensional hydrodynamic simulations~\cite{Ugliano:2012fvp, Sukhbold:2015wba, Horiuchi:2017qja, Suwa:2019svl, Vartanyan:2019ssu, Bollig:2020phc, Vartanyan:2023zlb}, which can differ markedly in their treatment of the successive emission phases, especially for the heavy-flavor neutrinos. We adopt one representative of each: the analytic model of Ref.~\cite{pagliaroli2009improved} (hereafter Pagliaroli09) and the 2D simulation of Ref.~\cite{Vartanyan:2023zlb} (hereafter Vartanyan23).  Within a Bayesian framework, we fit their predicted spectra to the SN1987A events to constrain the scalar coupling $\lambda_{\phi\nu}$ and mediator mass $m_\phi$. We treat the coupling diagonal in the neutrino mass basis, and analyze a mass-independent coupling and a mass-proportional one. The resulting limits depend on the emission model, the treatment of neutrino flavor, and the coupling structure. 
This shows that new-physics limits from SN1987A cannot be quoted independently of the flavor and heavy-flavor modeling or the coupling structure. This is due to the limited statistics of SN1987A. We then show a future high-statistics burst, such as one recorded by Hyper-Kamiokande, could give a meaningful upper bound on $\lambda_{\phi\nu}$.

This paper is organized as follows. In section~\ref{sec3:resonance} we introduce the effective interaction and calculate the differential flux depletion of the SN neutrino spectra from the resonant production of the scalar during propagation. In section~\ref{sec:n_osc} we introduce the two SN neutrino emission models and discuss how flavor conversion shapes the detected flux. We construct the Bayesian framework for parameter estimation in section~\ref{sec:bayesian} and present the results and discussion in section~\ref{sec:results}. Conclusions are drawn in section~\ref{sec:conclusions}. Technical details of the seesaw constructions, the SN neutrino flux models, the detection, and the statistical analysis are collected in the appendices.

%{\bf Introduction}---%
%
\section{Resonant Annihilation and Flux Attenuation}\label{sec3:resonance}

We consider the low-energy effective interactions
\begin{equation}\label{lagrangian}
  \mathcal{L} \supset -\dfrac{1}{2}\lambda^{ij}_{\phi \nu}\, \phi \nu_i \nu_j - \dfrac{1}{2} \lambda_{\phi\chi}\, \phi \chi \chi  + h.c. \ ,
\end{equation}
where $\nu_i$ are the neutrino mass eigenstates and $\chi$ is a massless dark fermion. The dark coupling may be as large as $\lambda_{\phi\chi}\sim\mathcal{O}(1)$, while the neutrino coupling $\lambda_{\phi\nu}$ is the quantity we aim to constrain. This interaction can descend from UV completed models such as  a Type-I seesaw or an inverse seesaw, both of which give rise to the coupling matrix $\lambda_{\phi\nu}^{ij}$ \emph{diagonal in the mass basis} of the active neutrinos.  As shown in Appendix~\ref{ap:seesaw}, the two models differ in how $\lambda_{\phi\nu}$ depends on the eigenstate mass $m_i$. The inverse seesaw could give the same coupling for all eigenstates, whereas the Type-I seesaw gives one proportional to the mass,
\begin{equation}\label{eq:coupling_structures}
\lambda_{\phi\nu}=
\begin{cases}
\lambda_0, & \text{inverse seesaw},\\[4pt]
\lambda_0\,m_i/T_\nu, & \text{Type-I seesaw}.
\end{cases}
\end{equation}
We choose this parametrization so that $\lambda_0$ is the variable fitted directly.

As the SN neutrinos propagate to Earth they traverse the cosmic neutrino background (C$\nu$B). Through the interaction Eq.~\eqref{lagrangian}, an SN antineutrino can annihilate resonantly with a relic neutrino into a pair of dark fermions, $\bar\nu(k)\,\nu(p)\to\phi\to\chi\bar\chi$. Since $\lambda_{\phi\nu}\ll1$, we neglect the regeneration of neutrinos from the dark radiation. This process thus depletes the SN neutrino flux. With $E_\nu$ the energy of the propagating SN neutrino, the differential flux $\Phi_i(E_\nu)$ of the $i$-th mass eigenstate obeys
\begin{equation}\label{eq:dPhi}
\frac{d\Phi_i}{dt}=-\Gamma_i(E_\nu)\,\Phi_i~.
\end{equation}
The attenuation rate $\Gamma_i$ is the thermal averaged interaction rate with the relic neutrino~\cite{Gondolo:1990dk,Weiler:1982qy},
\begin{equation}\label{eq:rate}
\Gamma(E_\nu)=\int\frac{d^3q}{(2\pi)^3}\,f_{\CnuB}(|\vec q\,|)\,\sigma\,v_{\rm M\o l},
\qquad f_{\CnuB}(|\vec q\,|)=\frac{1}{e^{|\vec q\,|/T_\nu}+1},
\end{equation}
where $T_\nu=1.68\times10^{-4}\,$eV is the C$\nu$B temperature today~\cite{Lesgourgues:2006nd,Long:2014zva}. 
For the $s$-channel process with $s=(k+p)^2$, the annihilation cross section derived in Appendix~\ref{ap:xsec} reads
\begin{equation}\label{eq:sigma}
\sigma(s)=\frac{\lambda_{\phi\nu}^2\lambda_{\phi\chi}^2}{16\pi}\,
\frac{\sqrt{s(s-4m_\nu^2)}}{(s-m_\phi^2)^2+m_\phi^2\Gamma_\phi^2},
\qquad
\Gamma_\phi\simeq\frac{\lambda_{\phi\chi}^2m_\phi}{16\pi}~.
\end{equation}
Since the SN neutrino is ultra-relativistic, i.e. $E_\nu\gg m_\nu$, we set $k\simeq E_\nu(1,0,0,1)$. Since at least two out of the three relic neutrino species are non-relativistic, $p=(E_1,p_1\sin\theta,0,p_1\cos\theta)$, where $E_1, p_1$ are the relic neutrino's energy and momentum, and $\theta$ is the collision angle. Thus,  $s\simeq2E_1E_\nu(1-\beta\cos\theta)$ with $\beta\equiv p_1/E_1$, and
\begin{equation}\label{eq:sigmav}
\sigma\,v_{\rm M\o l}\simeq\frac{\lambda_{\phi\nu}^2\lambda_{\phi\chi}^2}{8\pi}\,
\frac{E_1E_\nu(1-\beta\cos\theta)^2}{(s-m_\phi^2)^2+m_\phi^2\Gamma_\phi^2}\simeq \frac{\lambda_{\phi\nu}^2}{2 E_1E_\nu}\,
\frac{s^2\Gamma_\phi/m_\phi}{(s-m_\phi^2)^2+m_\phi^2\Gamma_\phi^2}.
\end{equation}

Finally for the $i$-th mass eigenstate Eq.~\eqref{eq:dPhi} integrates to $\Phi_i(E_\nu)=\Phi_i^{0}(E_\nu)\,e^{-\tau_i(E_\nu)}$, where 
\begin{equation}\label{eq:flux_1}
\tau_i(E_\nu)=\int_0^{d_{\rm SN}}\Gamma_i(E_\nu)\,dt\simeq\Gamma_i(E_\nu)\,d_{\rm SN},
\end{equation}
with $d_{\rm SN}\simeq 50\,$kpc the distance to SN1987A. Here $\Phi_i^0(E_\nu)$ is the flux reaching Earth in the absence of scattering.

\subsection{Resonance in the non-relativistic and relativistic limits}
For the non-relativistic relics, i.e. $\beta\to 0$, $E_1\to m_\nu$, the angular dependence drops out and we get 
\begin{equation}
\sigma\,v_{\rm M\o l}\simeq\frac{\lambda_{\phi\nu}^2\lambda_{\phi\chi}^2}{8\pi}\,
\frac{m_\nu E_\nu}{(2E_\nu m_\nu-m_\phi^2)^2+m_\phi^2\Gamma_\phi^2},
\end{equation}
which peaks at $E_\nu=m_\phi^2/(2m_\nu)$. The range of SN-neutrino energies over which the scattering is significantly enhanced is
\begin{equation}\label{eq:dEk}
\delta E_\nu=\frac{m_\phi\Gamma_\phi}{2m_\nu}=\frac{\lambda_{\phi\chi}^2m_\phi^2}{32\pi m_\nu}
\simeq{\rm keV}\left(\frac{10^{-1}\,{\rm eV}}{m_\nu}\right)\left(\frac{\lambda_{\phi\chi}}{1}\right)^2\left(\frac{m_\phi}{100\,{\rm eV}}\right)^2 .
\end{equation}
In this case, the scattering rate has a simple expression when on resonance: 
\begin{equation}
\Gamma^{\rm res}_{\rm non}(E_\nu)\approx\frac{8\pi \lambda_{\phi\nu}^2}{\lambda_{\phi\chi}^2E_\nu m_\nu}n_{\CnuB},\quad n_{\CnuB}=
\frac{3T^3_\nu \zeta (3) }{4\pi^2}
\end{equation}

For relativistic relics, i.e. $\beta\to1$, $E_1\gg m_\nu$, we now have $s=4E_1E_\nu\sin^2(\theta/2)$, and the resonance condition $4E_1E_\nu\sin^2(\theta/2)=m_\phi^2$ depends on the collision angle. Therefore, at a fixed $E_\nu$, it probes a larger range of $m_\phi$ than a non-relativistic one.

\subsection{Attenuation rate and flux depletion}

Carrying out the phase-space integral Eq.~\eqref{eq:rate} with the kernel Eq.~\eqref{eq:sigmav} gives
\begin{equation}\label{eq:scattering_rate}
\Gamma(E_\nu)=\frac{T_\nu^3\lambda_{\phi\nu}^2}{4\pi^2m_\phi^2}\,\mathcal{I}(m_\nu,m_\phi,E_\nu),
\end{equation}
with the dimensionless function
\begin{equation}\label{eq:normalized_rate}
\mathcal{I}=\int_{0}^\infty  dx\,\frac{x^2/\sqrt{x^2+x_0^2}}{e^{x}+1}\int_{-1}^{1}dy\;
\frac{w\alpha\left(\sqrt{x^2+x_0^2}-x y\right)^2}
{\left(\alpha\left( \sqrt{x^2+x_0^2}-x y\right)-1\right)^2+w^2},
\end{equation}
where $x=p_1/T_\nu$, $x_0=m_\nu/T_\nu$, $\alpha=2E_\nu T_\nu/m_\phi^2$, and $w=\Gamma_\phi/m_\phi=\lambda_{\phi\chi}^2/16\pi$.

\begin{figure}[th]
    \centering    
    \includegraphics[width=1\linewidth]{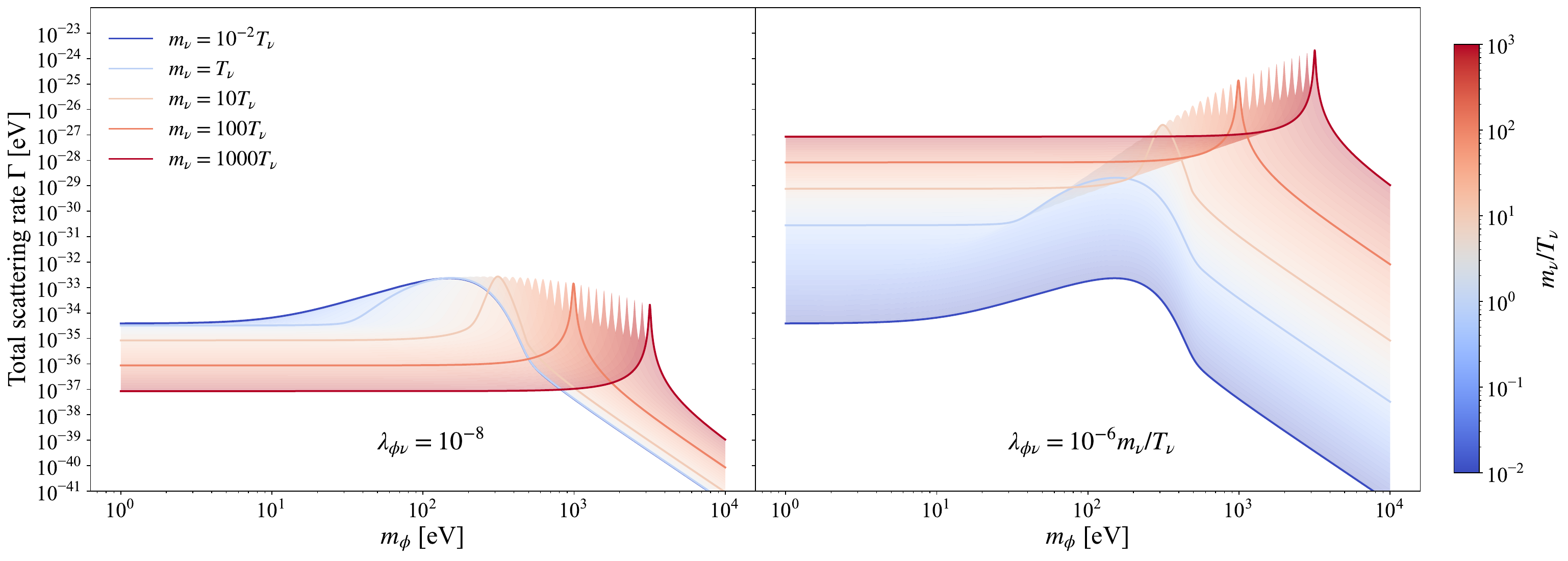}
    \caption{Scattering rate $\Gamma$ (Eq.~\eqref{eq:scattering_rate}) as a function of $m_\phi$, at fixed SN-neutrino energy $E_\nu=30\,$MeV and dark coupling $\lambda_{\phi\chi}=1$. The curves span relic neutrino masses from relativistic, $m_\nu=10^{-2}T_\nu$ in blue, to non-relativistic, $m_\nu=10^{3}T_\nu$ in red, as indicated by the color bar. The two panels correspond to the coupling structures of Eq.~\eqref{eq:coupling_structures}. The left panel uses the mass-independent coupling $\lambda_{\phi\nu}=10^{-8}$. The right panel uses the mass-proportional coupling $\lambda_{\phi\nu}=10^{-6}\,m_\nu/T_\nu$, for which the rate is enhanced by $(m_\nu/T_\nu)^2$ so that the ordering of the curves inverts and the heaviest relics dominate.}
    \label{fig:norm_rate}
\end{figure}

Fig.~\ref{fig:norm_rate} shows $\Gamma$ as a function of $m_\phi$ at fixed $E_\nu=30\,$MeV and $\lambda_{\phi\chi}=1$ for several relic masses, comparing the mass-independent coupling in the left panel with the mass-proportional one in the right panel. The kinematic factor $\mathcal{I}$ is larger for relativistic relics, since the collision angle provides an extra handle for meeting the resonance condition. With the mass-proportional coupling the $(m_\nu/T_\nu)^2$ enhancement overcomes this kinematic preference and inverts the ordering, so the heaviest relics dominate. The $m_\phi$ dependence is common to both panels. Below resonance, as $m_\phi$ decreases and $\alpha$ grows correspondingly, Eq.~\eqref{eq:normalized_rate} scales as $\propto m_\phi^2$, cancelling the $1/m_\phi^2$ prefactor so that $\Gamma$ becomes independent of $m_\phi$ and produces the left plateau. Above resonance, $\Gamma\propto m_\phi^{-4}$.

To estimate the depletion near resonance, we take a relativistic relic $E_1\sim T_\nu$ and the kinematic region $\alpha\simeq1/2$, corresponding to $m_\phi^2\simeq4E_\nu T_\nu$. With $\mathcal{I}\sim\mathcal{O}(1)$,
\begin{equation}
\tau(E_\nu)=\Gamma(E_\nu)d_{\rm SN}\approx\frac{\lambda_{\phi\nu}^2T_\nu^2d_{\rm SN}}{16\pi^2E_\nu}
\sim0.1\left(\frac{\lambda_{\phi\nu}}{10^{-6}}\right)^2\left(\frac{10\,{\rm MeV}}{E_\nu}\right),
\end{equation}
and the associated mediator mass is $m_\phi\simeq2\sqrt{T_\nu E_\nu}\simeq82\,{\rm eV}\,(E_\nu/10\,{\rm MeV})^{1/2}$.

\section{The Detected SN Neutrino Flux}\label{sec:n_osc}

To confront the resonant attenuation of Sec.~\ref{sec3:resonance} with the SN1987A data, we need the electron-antineutrino flux that reaches the detector. Two ingredients enter, a model of the neutrino emission at the source, and the flavor conversion the neutrinos undergo on their way out of the star. We first introduce the two emission models adopted, and then describe the flavor conversion, assembling the detected $\bar\nu_e$ flux including the propagation attenuation which enters the statistical analysis.

\subsection{SN neutrino emission models}\label{sec:emission_models}

Predicting the detected signal requires the SN-neutrino flux at the source. We employ two complementary descriptions and compare them throughout: a parameterized, bottom-up model, Pagliaroli09~\cite{pagliaroli2009improved}, and a first-principles 2D radiation-hydrodynamic simulation, Vartanyan23~\cite{Vartanyan:2023zlb}. The detailed flux parameterizations are collected in Appendix~\ref{ap:flux}. Here we summarize their main features.

Pagliaroli09 describes the neutrino-driven emission through two stages. A brief but luminous accretion phase, powered by positron capture on free neutrons ($e^+ n\to p\,\bar\nu_e$), is followed by a longer cooling phase that carries most of the emitted energy and radiates all flavors with comparable luminosities. The emission is controlled by six astrophysical parameters: the neutrinosphere radius $R_c$, the cooling temperature $T_c$ and timescale $\tau_c$, and the accreting mass $M_a$, temperature $T_a$, and timescale $\tau_a$. Crucially, because accretion-phase emission is dominated by charged-current capture, which preferentially produces electron flavors, the heavy-flavor antineutrino flux $\bar\nu_x$ is taken to be negligible during accretion and switched on only in the cooling phase.

Vartanyan23 instead provides fixed neutrino spectra obtained from \texttt{FORNAX} radiation-hydrodynamic simulations of core-collapse supernovae, with no adjustable source parameters once a progenitor is selected. Because the SN1987A progenitor is expected in the $15$--$20\,M_\odot$ range, we use the 22 simulated spectra falling in this window. In these simulations the heavy-flavor antineutrinos are produced through neutral-current pair processes from the earliest times, so a sizable $\bar\nu_x$ flux is present already during the accretion phase, comparable in magnitude to $\bar\nu_e$.

\subsection{Flavor conversion and the detected flux}

The scattering rate $\Gamma$ is formulated in the neutrino mass basis, whereas the SN emission models are specified in the flavor basis, so the flavor states produced near the neutrinosphere must be mapped onto the vacuum mass eigenstates that propagate outward. Since the detectors are sensitive to electron antineutrinos, we focus on the antineutrino channel and assume \emph{normal ordering} where $\nu_1$ is the lightest neutrino.

Deep inside the SN the matter potential dominates the propagation Hamiltonian, so the emitted flavor states coincide with the instantaneous matter eigenstates. For antineutrinos in the normal ordering, as shown in Appendix~\ref{ap:flavor}, these connect adiabatically to the vacuum mass eigenstates $\bar\nu_{i}$ as
\begin{equation}\label{eq:matter_id}
\bar\nu_e = \bar\nu_{1}, \quad \bar\nu_{\mu} = \bar\nu_{2}, \quad \bar\nu_{\tau} = \bar\nu_{3}~.
\end{equation}
We denote by $\Phi_i^0$ the flux of the $i-$th antineutrino mass eigenstate reaching Earth in the absence of scattering. The electron-antineutrino flux is then $\Phi_1^0=\Phi_{\bar\nu_e}^0$, while the two heavy flavors, treated equally in SN emission models, give $\Phi_2^0=\Phi_3^0=\Phi_{\bar\nu_x}^0$. 
Since the detectors record only electron antineutrinos, the detected $\bar\nu_e$ flux is thus given by
\begin{equation}\label{eq:mixing_flux_detector}
    \Phi_{\bar\nu_e}^{\rm osc}(m_\phi,\lambda_{\phi\nu}) =  |U_{e1}|^2e^{-\tau_1}\Phi_{\bar\nu_e}^0 +\left(|U_{e2}|^2e^{-\tau_2} +|U_{e3}|^2e^{-\tau_3} \right)\Phi_{\bar\nu_x}^0~,
    %(1-|U_{e1}|^2)\Phi_{\bar\nu_x}^0 \ .
\end{equation}
where $U_{e1}, U_{e2}, U_{e3}$ are the corresponding elements from the PMNS matrix.  

Here, $\Phi_{\bar{\nu}_e}^0$ and $\Phi_{\bar{\nu}_x}^0$ are predicted by the two SN emission models detailed in Appendix~\ref{ap:flux} and displayed in Fig.~\ref{fig:flux_osc_a}. Also shown is the detected $\bar{\nu}_e$ flux in the absence of scattering (i.e., by setting $\lnu = 0$).
Because Pagliaroli09 emits almost no heavy flavor during the accretion phase, its detected flux $\Phi_{\bar\nu_e}^{\rm osc}$ is dominated by the electron-antineutrino term $|U_{e1}|^2\Phi_{\bar\nu_e}^0$, whereas in Vartanyan23 the electron and heavy-flavor components are comparable throughout.

\begin{figure*}
    \centering
    \subfigure[]{
        \includegraphics[width=0.3\textwidth]{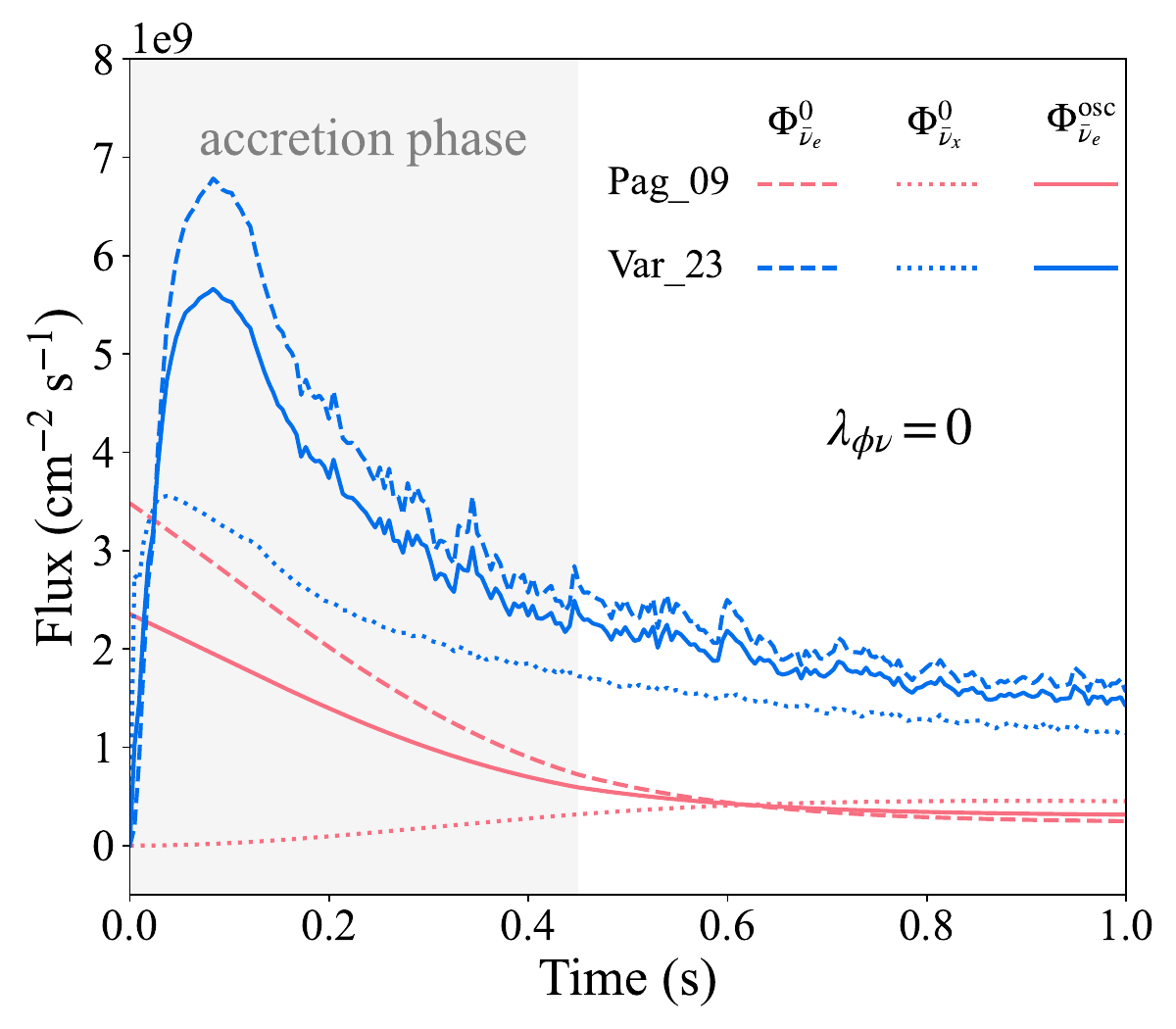}
        \label{fig:flux_osc_a}
    }
    % \hfill
    \subfigure[]{
        \includegraphics[width=0.3\textwidth]{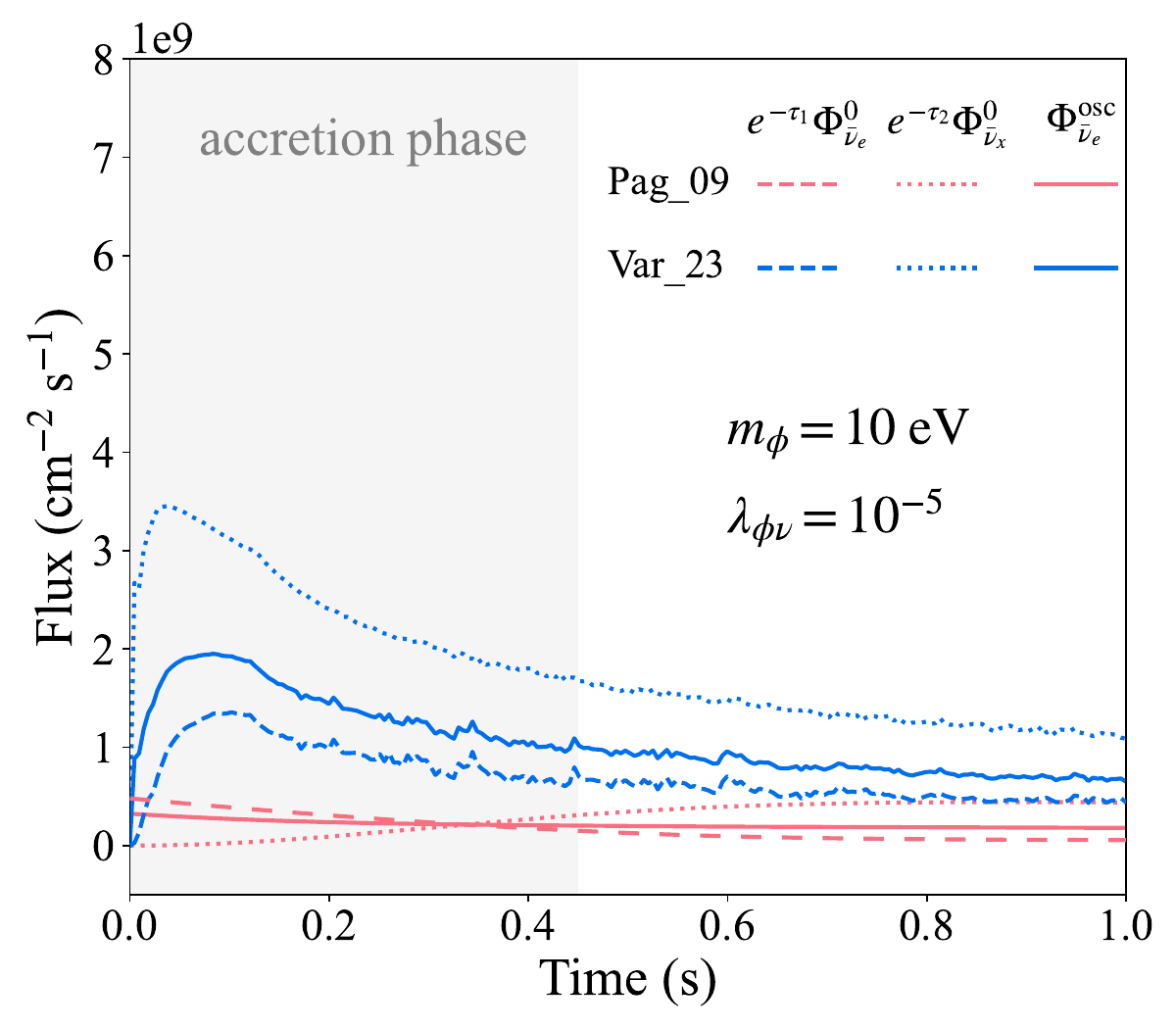}
        \label{fig:flux_osc_b}
    }
    % \hfill
    \subfigure[]{
        \includegraphics[width=0.3\textwidth]{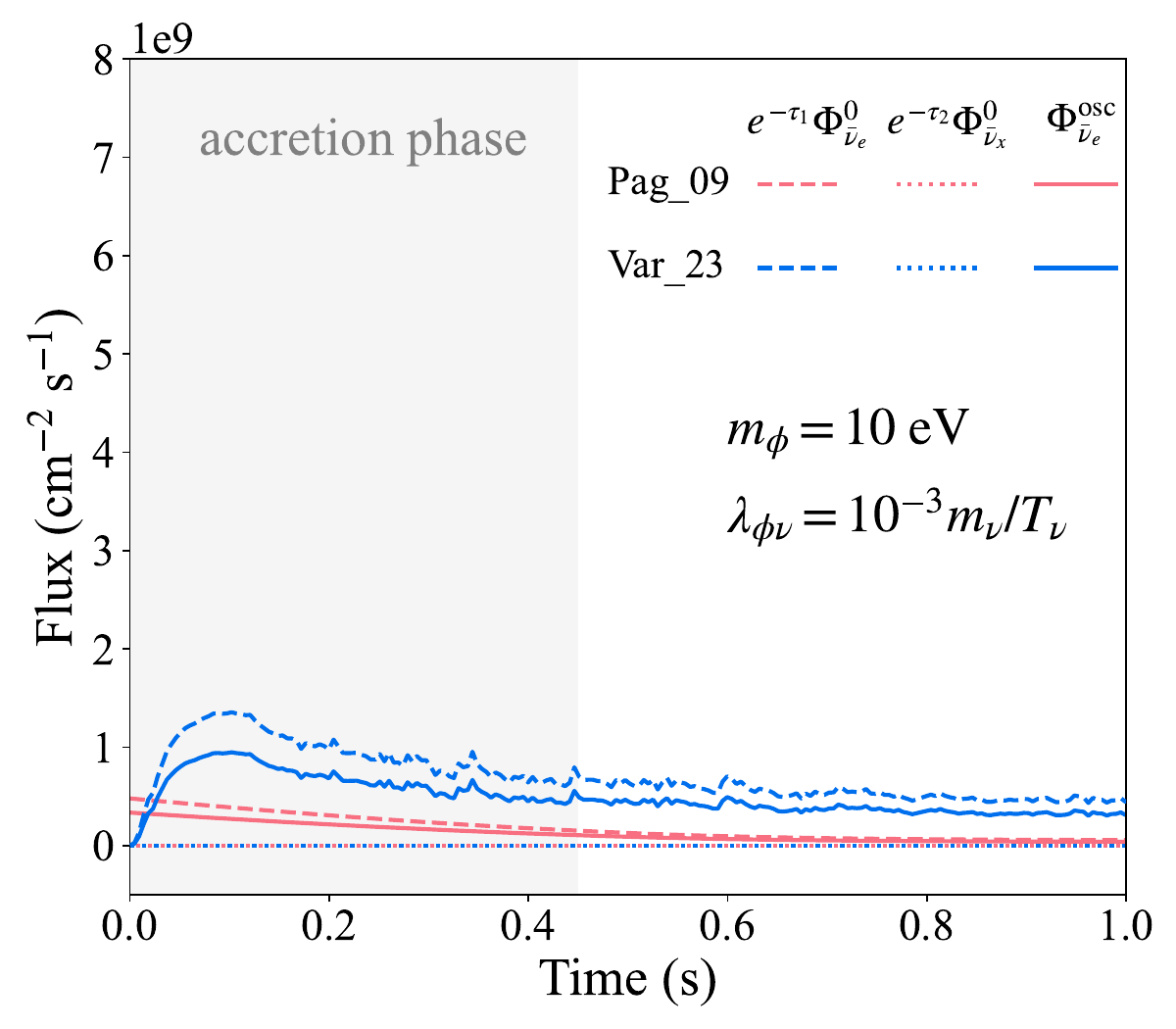}
        \label{fig:flux_osc_c}
    }
    \caption{Antineutrino fluxes on Earth versus time, for Pagliaroli09 (red) and Vartanyan23 (blue). Panel (a) is for $\lambda_{\phi\nu}=0$ without scattering. Panels (b) and (c) include scattering off the C$\nu$B at $m_\phi=10\,$eV, for the mass-independent coupling $\lambda_{\phi\nu}=10^{-5}$ in (b) and the mass-proportional coupling $\lambda_{\phi\nu}=10^{-3}\,m_\nu/T_\nu$ in (c). The dashed (dotted) lines show the electron (heavy-flavor) antineutrino fluxes. The solid lines show the detected flux after taking into account of oscillation $\Phi_{\bar\nu_e}^{\rm osc}$ (Eq.~\eqref{eq:mixing_flux_detector}).  Pagliaroli09 is evaluated with the parameter set $R_c = 10\rm~km$, $T_c = 4.5\rm ~MeV$, $\tau_c = 4.5\rm~s$, $M_a =0.1M_\odot$, $T_a = 2 \rm~MeV$, and $\tau_a = 0.5\rm~s$, and  Vartanyan23 with the $M=15.01\,M_\odot$ spectrum. The gray band marks the accretion phase, cut at $\tau_a=0.45\,$s. Here we asuume normal mass ordering and set the lightest neutrino mass $m_1 = 10^{-2}~T_\nu$. The PMNS mixing parameters and mass-squared differences are fixed to the best-fit values from the latest global analysis (including SK atmospheric data)~\cite{Esteban:2024eli}: $|U_{e1}|^2 = 0.677$, $|U_{e2}|^2 = 0.301$, $|U_{e3}|^2 = 0.022$, $\Delta m_{21}^2 = 7.49 \times 10^{-5}\text{ eV}^2$, and $\Delta m_{31}^2 = 2.513 \times 10^{-3}\text{ eV}^2$.}
    \label{fig:flux_osc}
\end{figure*}

In Fig.~\ref{fig:flux_osc_b} and Fig.~\ref{fig:flux_osc_c}, we include the effect of flux attenuation due to $\bar\nu\nu$ scattering assuming two different types of $\lambda_{\phi\nu}$. For the mass-independent coupling in Fig.~\ref{fig:flux_osc_b}, the heavy mass eigenstates $\bar\nu_2$ and $\bar\nu_3$ are non-relativistic, and hence their scattering rates are kinematically suppressed, leading to $\tau_2,\tau_3\ll\tau_1$. Only the $\bar\nu_1$ component, detected as $\bar\nu_e$, is then appreciably attenuated. 
For the mass-proportional coupling in Fig.~\ref{fig:flux_osc_c}, the pattern is reversed. Because the coupling scales with neutrino mass, the heavy eigenstates couple far more strongly than $\bar{\nu}_1$ and become heavily depleted. Furthermore, since $m_2 \neq m_3$, the two heavy eigenstates are attenuated by different amounts.
In Fig.~\ref{fig:flux_osc} we only show $e^{-\tau_2} \Phi_{\bar\nu_x}^0$ as a representation for the attenuated fluxes.  $e^{-\tau_3} \Phi_{\bar\nu_x}^0$ has a similar pattern. 

Finally, for comparison we define a scenario without taking into account of flavor conversion. We treat the flavor blind detected flux as
\begin{equation}\label{eq:flux_detector}
\Phi_{\bar{\nu}_e}^{\text{f-blind}} = \Phi_{\bar{\nu}_e}^0 \times e^{-\tau_1} 
\end{equation}
in the rest of the paper.

\section{Statistical Framework}\label{sec:bayesian}

Section~\ref{sec:n_osc} gives the detected flux $\Phi_{\bar\nu_e}^{\rm osc}$ (Eq.~\eqref{eq:mixing_flux_detector}) for each emission model. We now confront it with the SN1987A data by constructing a per-event likelihood and estimating the model parameters. The data are the $\bar\nu_e$ events recorded by Kamiokande-II, IMB, and Baksan~\cite{Kamiokande-II:1987idp, Kamiokande-II:1989hkh, bionta1987observation,bratton1988angular, Alekseev:1988gp}. After the standard event selection, detailed in Appendix~\ref{ap:detection} and Table~\ref{tab:events}, 24 events enter the Pagliaroli09 analysis over a $30\,\mathrm{s}$ window, whereas the Vartanyan23 analysis is restricted to the first $4\,\mathrm{s}$ covered by the simulations, leaving 17 events.
 
In all three detectors, the signal arises from positrons produced via inverse beta decay (IBD), $\bar{\nu}_e p\to e^+n$.
Each event is characterized by three observables $(t, E_e, c_\theta \equiv \cos\theta)$, the event time $t$, the positron energy $E_e$, and the scattering angle $\theta$ between the outgoing $e^+$ and the incoming $\bar{\nu}_e$. 
The differential signal rate of positrons at time $t$ combines the detected flux with the IBD cross section $\sigma$, the number of target protons $N_p$, the angular bias $\xi(c_\theta)$, and the energy-dependent efficiency $\eta(E_e)$,
\begin{equation}\label{eq:signalrate}
\frac{d S}{dE_{e}dc_\theta}(t,E_e,c_\theta)=  N_p \times\frac{d\Phi}{dE_\nu}(t,E_\nu)\times \frac{d\sigma}{d c_\theta}(E_\nu,c_\theta)  \times
\frac{dE_\nu}{dE_e}\times \xi(c_\theta)\times \eta(E_e) \ ,
\end{equation}
where $E_\nu$ is the antineutrino energy and $\Phi$ is the detected $\bar\nu_e$ flux from either emission model.

We adopt an unbinned extended Poisson likelihood for the events observed in each detector,
\begin{equation}\label{eq:likelihood}
\mathcal{L}_d \propto e^{-N_{\text{exp}}} \prod_{i=1}^{N_d} \mathcal{R}(t_i, E_i, c_{\theta, i}),
\end{equation}
where $N_{\text{exp}}$ is the total expected number of signal events over the observation window and $\mathcal{R}(t_i, E_i, c_{\theta, i})$ is the theoretical differential event rate at the measured time, positron energy, and scattering angle of the $i$-th event. In contrast to the bare signal rate $S$ of Eq.~\eqref{eq:signalrate}, $\mathcal{R}$ additionally incorporates the background contamination, the energy-resolution smearing, and the live-time fraction. Its full detector-specific form is given in Appendix~\ref{ap:detection}.

Parameter estimation proceeds from the posterior distribution given by Bayes' theorem,
\begin{equation}
P(\boldsymbol{\Theta} | \text{Events}) = \mathcal{L}_d(\text{Events} | \boldsymbol{\Theta})\,P_0(\boldsymbol{\Theta}),
\end{equation}
with $\boldsymbol\Theta$ the parameter space and $P_0(\boldsymbol\Theta)$ the priors. For Pagliaroli09 the emission is described by six astrophysical parameters and the two new-physics parameters $(m_\phi,\lambda_0)$, so that $\boldsymbol\Theta_{\rm P} = [R_c, T_c, \tau_c, M_a, T_a, \tau_a, m_\phi, \lambda_0]$. For Vartanyan23 the spectra are fixed once a progenitor is chosen, so the parameter space reduces to the two new-physics parameters, $\boldsymbol\Theta_{\rm v} = [m_\phi, \lambda_0]$. We take uniform priors on all parameters, listed in Tables~\ref{tab:best_fit} of Appendix~\ref{ap:stats}.

Throughout the analysis we assume normal ordering and fix the lightest neutrino mass to $m_1=10^{-2}T_\nu$ and the dark-sector coupling to $\lambda_{\phi\chi}=1$. The neutrino mixing-matrix elements and mass-squared splittings are fixed to the latest global-fit values~\cite{Esteban:2024eli}, quoted in the caption of Fig.~\ref{fig:flux_osc}. The posterior is sampled with a Markov-chain Monte Carlo (MCMC) for the SN1987A data. The details of this numerical analysis are given in Appendices~\ref{ap:detection} and~\ref{ap:stats}.

To disentangle the effects of neutrino flavor conversion and of the coupling structure, we analyze three benchmark scenarios. Case I is the flavor-blind reference, in which oscillations are neglected and the detected flux is the attenuated electron-antineutrino flux of Eq.~\eqref{eq:flux_detector}. Case II includes flavor conversion with the mass-independent  coupling of Eq.~\eqref{eq:coupling_structures}. Case III includes flavor conversion with the mass-proportional coupling of Eq.~\eqref{eq:coupling_structures}.

%%%%%%%%%%%%%%%%%%%%%%%%%%%%%%%%%%%%%%
\section{Results and Discussion}\label{sec:results}

For each scenario we analyze both emission models. The marginalized posteriors and $1\sigma$, $2\sigma$, and $3\sigma$ highest-posterior-density (HPD) credible regions for the new-physics parameters are shown in Fig.~\ref{fig:corner_plots}.\footnote{Throughout this work, we use the terms $1\sigma$, $2\sigma$, and $3\sigma$ as linguistic shorthand to denote the $68\%$, $95\%$, and $99.7\%$ credible levels, respectively, regardless of whether we refer to 1D or 2D posterior distributions.} The full Pagliaroli09 posteriors and best-fit values are given in Figs.~\ref{fig:8p_I}--\ref{fig:8p_III} and Table~\ref{tab:best_fit}. The inferred six emission parameters remain consistent with Ref.~\cite{pagliaroli2009improved}, so the new interaction does not bias the source properties. For Vartanyan23, each of the 22 progenitor spectra is fit independently (Figs.~\ref{fig:2d_sim_all_I}--\ref{fig:2d_sim_all_III}), and the resulting chains are stacked with equal weight to give a conservative bound.

\subsection{Case-by-case results}

\begin{figure*}
    \centering
    \subfigure{
        \includegraphics[width=0.3\textwidth]{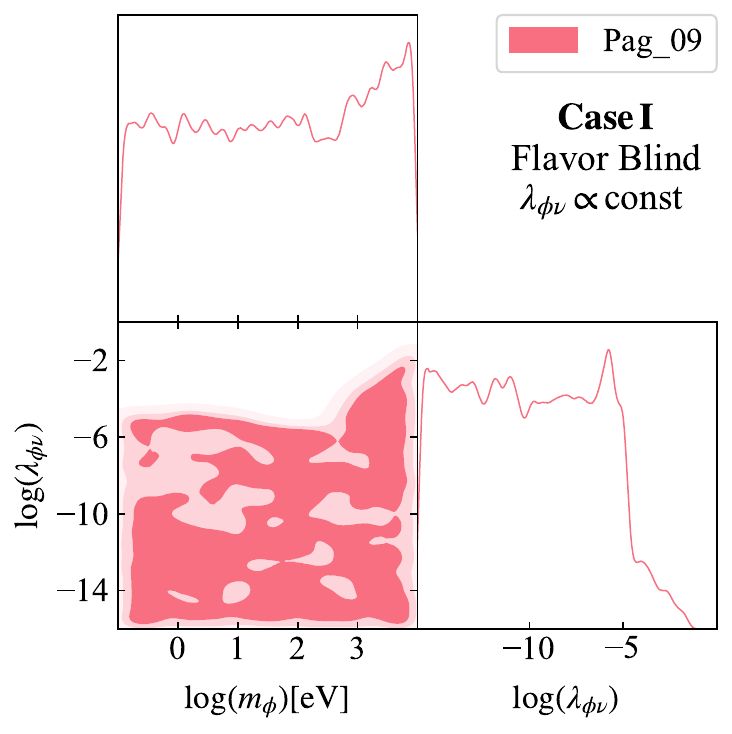}
        \label{fig:P_corner_I}
    }
    \hfill
    \subfigure{
        \includegraphics[width=0.3\textwidth]{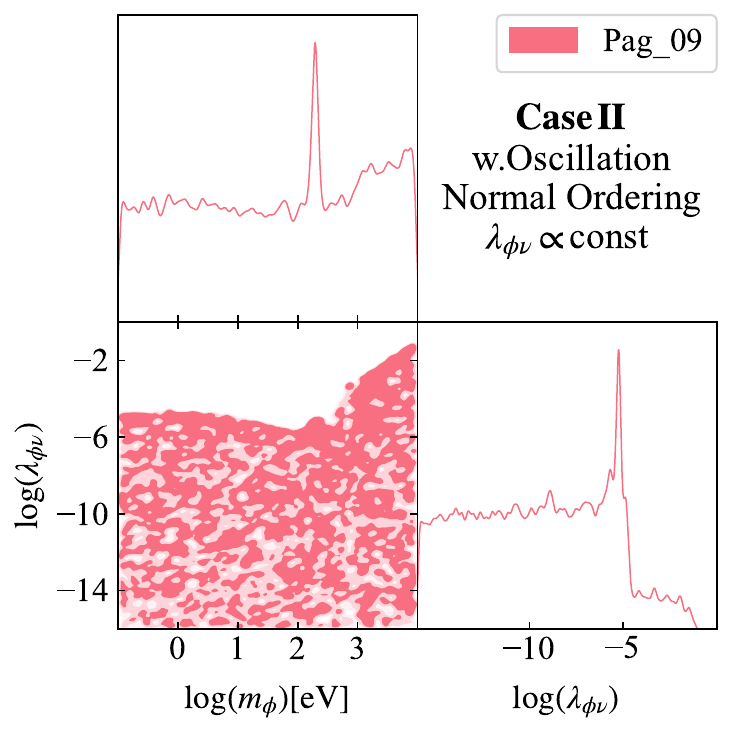}
        \label{fig:P_corner_II}
    }
    \hfill
    \subfigure{
        \includegraphics[width=0.3\textwidth]{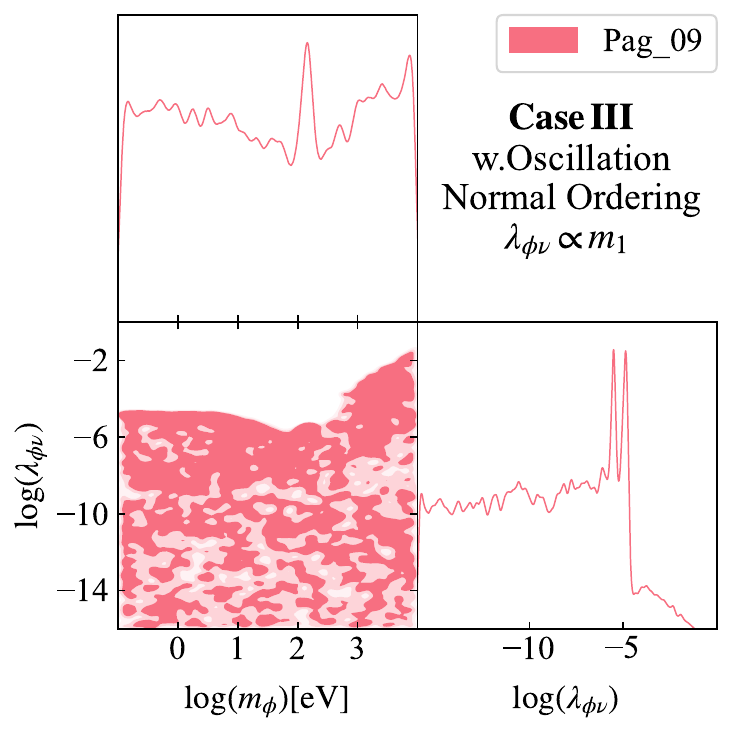}
        \label{fig:P_corner_III}
    }
    
    \subfigure{
        \includegraphics[width=0.3\textwidth]{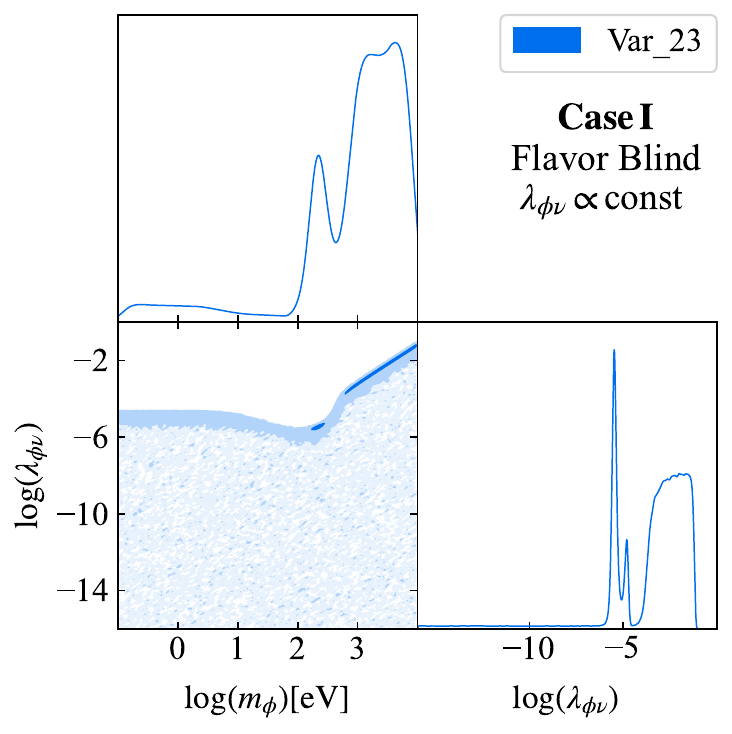}
        \label{fig:V_corner_I}
    }
    \hfill
    \subfigure{
        \includegraphics[width=0.3\textwidth]{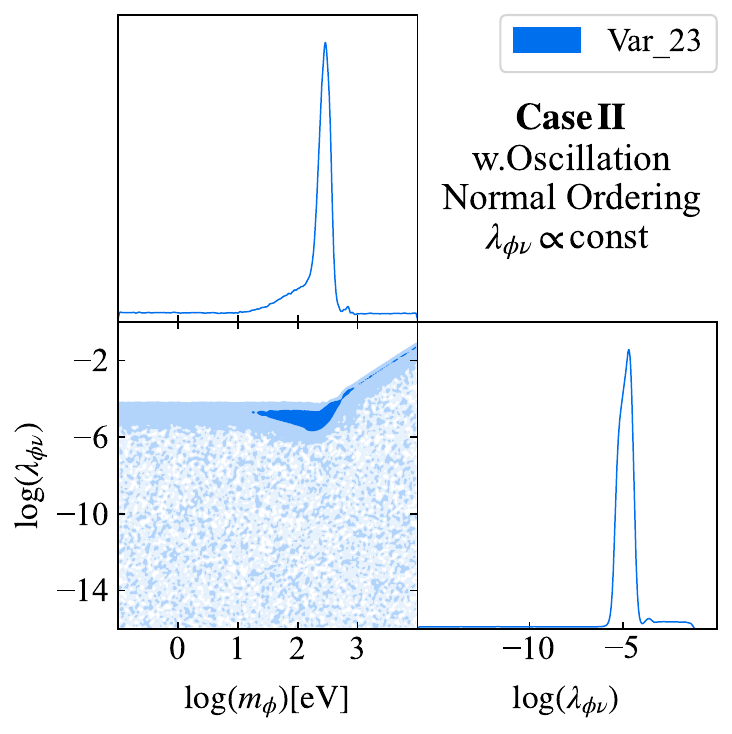}
        \label{fig:V_corner_II}
    }
    \hfill
    \subfigure{
    \includegraphics[width=0.3\textwidth]{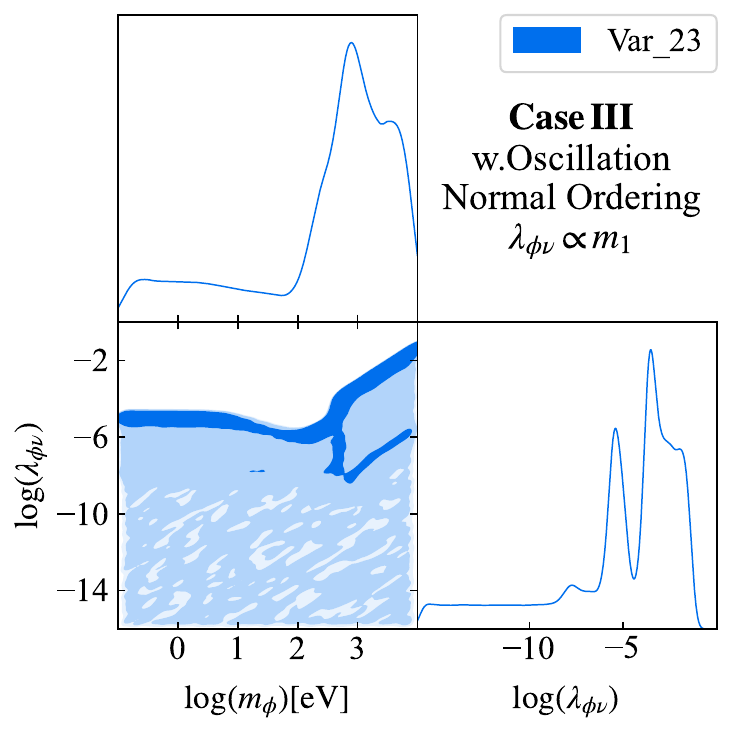}
    \label{fig:V_corner_III}
    }
    \caption{Joint 2D contours and 1D marginalized posteriors for the new-physics parameters $(m_\phi,\lambda_{\phi\nu})$ from the MCMC fits. The top panels use Pagliaroli09~\cite{pagliaroli2009improved}, obtained by fitting all eight parameters, and the bottom panels use Vartanyan23~\cite{Vartanyan:2023zlb}, obtained by stacking the 22 progenitor spectra. The three columns are Case I (flavor blind, left), Case II (oscillation with the mass-independent coupling, middle), and Case III (oscillation with the mass-proportional coupling, right), with $m_1 = 10^{-2}~T_\nu$. Shaded regions correspond to the $68\%$, $95\%$, and $99.7\%$ credible levels, from dark to light. }\label{fig:corner_plots}
\end{figure*}

In Case I the attenuation is treated as flavor blind, so the detected flux is the emitted $\bar\nu_e$ depleted by $e^{-\tau_1}$ of Eq.~\eqref{eq:flux_detector}. Pagliaroli09 shows no preference for a nonzero coupling and yields an upper bound on $\lambda_{\phi\nu}$, excluding the region of excessive depletion at the $2\sigma$ credible level (C.L.). Vartanyan23, whose predicted flux is somewhat higher, develops a mild best-fit preference for a nonzero coupling that attenuates the excess, but it still excludes large couplings at the $2\sigma$ level and its bound nearly coincides with that of Pagliaroli09. Because the depletion acts on the $\bar\nu_e$ flux in the same way for both models, the two emission models give closely matching bounds, as seen in the left panels of Fig.~\ref{fig:corner_plots}.

In Case II we include flavor conversion with the mass-independent coupling. The Pagliaroli09 bounds are essentially unchanged from Case I, whereas Vartanyan23 weakens substantially. With a mass-independent coupling the heavy eigenstates $\bar\nu_2,\bar\nu_3$ are non-relativistic and share the coupling of $\bar\nu_1$, so $\tau_2,\tau_3\ll\tau_1$ and the heavy-flavor term of Eq.~\eqref{eq:mixing_flux_detector} is barely depleted. Vartanyan23's large early $\bar\nu_x$ flux then supplies an essentially un-depleted contribution to the detected $\bar\nu_e$ channel that masks the depletion of $\bar\nu_1$, so even large couplings remain compatible with the data. Pagliaroli09 emits little heavy flavor during accretion, so its constraint is unaffected.

Besides, the different behavior of the MCMC chains is also related to the number of adjustable source parameters in each model. In Pagliaroli09, changes in the predicted spectrum induced by neutrino oscillations can be partially absorbed by varying the six SN-emission parameters. This flexibility allows the fit to accommodate changes in the overall normalization and spectral shape while retaining sensitivity to the energy-dependent depletion governed by $m_\phi$ and $\lambda_{\phi\nu}$. In contrast, the Vartanyan23 analysis is based on fixed simulated spectra with no parameters to adjust. Consequently, any oscillation-induced modification of the predicted spectrum is transferred more directly to the posterior distributions of the new-physics parameters. This effect is particularly relevant given the limited number of observed SN1987A events.

In Case III we include flavor conversion with the mass-proportional coupling, and the two emission models agree once more. The coupling now grows with the eigenstate mass, and with the plotted $\lambda_{\phi\nu}\propto m_1$, the heavy eigenstates carry couplings $\lambda_{\phi\nu}(m_{2,3})=\lambda_{\phi\nu}(m_1)\,m_{2,3}/m_1$ that are three to four orders of magnitude larger. The constraint is therefore set not by the weakly-coupled $\bar\nu_1$ but by the depletion of $\bar\nu_2$ and $\bar\nu_3$ that reach the detector through oscillation. The value of $\lambda_{\phi\nu}(m_1)$ at which this depletion becomes significant is fixed by the eigenstate masses and couplings, independent of the emission model. Since both models supply heavy flavor over their fit windows, both reach this common threshold and their contours approximately coincide.

\subsection{Comparison with existing constraints and projected sensitivity}

\begin{figure}
    \centering
    \includegraphics[width=0.65\linewidth]{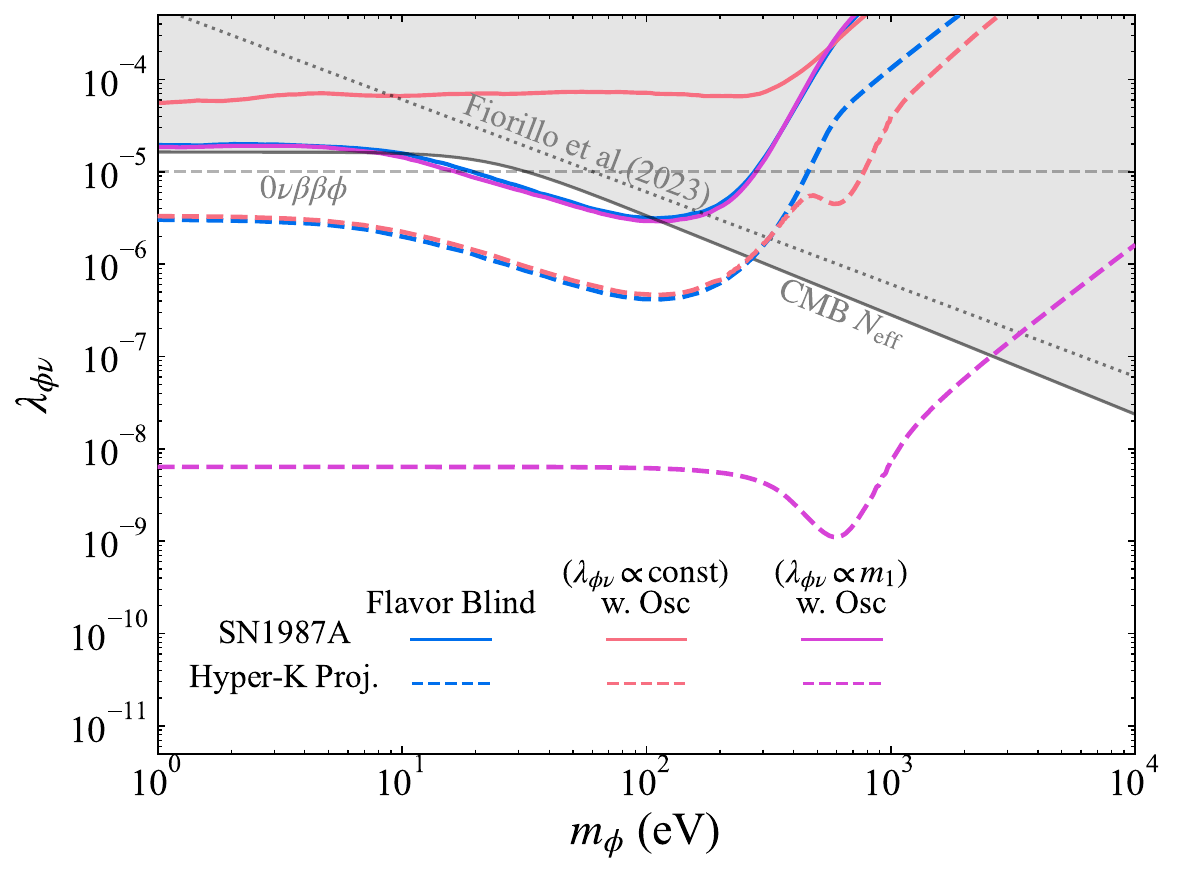}
    \caption{Constraints on $\lambda_{\phi\nu}$ as a function of $m_\phi$ for the Vartanyan23 SN model. The solid colored curves show the $2\sigma$ upper limits derived from the SN1987A data, while the dashed colored curves show the projected Hyper-Kamiokande sensitivity assuming ${\sim}2000$ events and the $M=15.01\,M_\odot$ spectrum. The colors correspond to the three scenarios: flavor-blind couplings (blue), flavor oscillations with mass-independent couplings (red), and flavor oscillations with mass-proportional couplings (magenta). For the last scenario, the constraint is shown in terms of $\lambda_{\phi\nu}(m_1)$. The gray curves indicate existing constraints from the cosmological bound $\Delta N_{\rm eff}<0.285$ (solid)~\cite{Li:2023puz}, searches for Majoron-emitting neutrinoless double-beta decay, $0\nu\beta\beta\phi$ (dashed)~\cite{KamLAND-Zen:2012uen}, and the nonobservation of ${\sim}100\,$MeV neutrino events from SN1987A (dotted)~\cite{Fiorillo:2022cdq}. Normal mass ordering is assumed with $m_1=10^{-2}T_\nu$.}
    \label{fig:contrast_plot}
\end{figure}

Figure~\ref{fig:contrast_plot} shows $2\sigma$ upper limits from the SN1987A data (solid lines) using Vartanyan23 model. %Solid lines are the $2\sigma$ upper limits from the SN1987A data and dashed lines the projected Hyper-Kamiokande sensitivity, shown for the 
Three scenarios are considered: the flavor-blind case in blue, the case including flavor conversion with the mass-independent coupling in red, and the case including flavor conversion with the mass-proportional coupling in magenta. For the mass-proportional couplings, we only show the coupling to the lightest mass eigenstate, $\lambda_{\phi\nu}(m_1)$, the constraints on $\lambda_{\phi\nu}(m_{2,3})$ can be obtained simply by rescaling.

With the current SN1987A data, Cases~I and~III give comparable limits. The bounds tighten from $\lambda_{\phi\nu}\sim10^{-5}$ at $m_\phi\sim1\,$eV to a few$\times10^{-6}$ around $m_\phi\sim100\,$eV, before weakening at larger masses as the resonant energy, $m_\phi^2/(2m_\nu)$, moves beyond the SN neutrino spectrum.
The limit in Case~II is markedly weaker because the unattenuated heavy-flavor component masks the depletion, preventing a pronounced resonant dip from developing and leaving the bound at approximately $\lambda_{\phi\nu}\sim10^{-4}$. Over much of the mass range considered, our limits are competitive with, and in some regions stronger than, those derived from the cosmological constraint $\Delta N_{\rm eff}<0.285$~\cite{Li:2023puz}, searches for Majoron-emitting neutrinoless double-beta decay, $0\nu\beta\beta\phi$~\cite{KamLAND-Zen:2012uen}, and the nonobservation of ${\sim}100\,$MeV neutrino events from SN1987A~\cite{Fiorillo:2022cdq}. The $0\nu\beta\beta\phi$ limit is reported at $90\%$ C.L., while all other limits shown, including ours, are given at $95\%$ C.L.

Finally, we forecast the sensitivity of next-generation detectors, taking Hyper-Kamiokande as our benchmark. Later and next-generation detectors, Super-Kamiokande~\cite{fukuda2003super_SK}, Hyper-Kamiokande~\cite{abe2018hyper_hk1, abe2021supernova_hk2}, DUNE~\cite{abi2021supernova_dune1, abed2023impact_dune2}, and JUNO~\cite{abusleme2024real_juno}, will record far more events than the small SN1987A sample. The larger fiducial mass enters our signal rate through the number of target protons, and, since Hyper-Kamiokande has not yet operated, we approximate its response by a $5\,$MeV threshold with $100\%$ efficiency above it. We evaluate the projected sensitivity with a profile-likelihood-ratio test. Following Wilks' theorem, the test statistic is
\begin{equation}\label{wilks}
    \chi^2 = -2\log  \frac{{\cal L}_d(m_\phi, \lambda_{\phi\nu})}{{\cal L} _d(m_\phi, \lambda_{\phi\nu} = 0)} .
\end{equation}
The denominator is the maximum likelihood under the null hypothesis of no new physics, and the numerator the likelihood with the new interaction. Both are evaluated using Eq.~\eqref{eq:likelihood} over the simulated events. At each fixed $m_\phi$ the likelihood ratio depends  on $\lambda_{\phi\nu}$ only, following a $\chi^2$ distribution with one degree of freedom. Therefore we set the $95\%$ C.L. upper limit at $\chi^2=3.84$. The projected limits for all three scenarios are shown as dashed lines in Fig.~\ref{fig:contrast_plot}, and clearly improve on the SN1987A limits by one to two orders of magnitude. In particular, the larger sample restores a bound in Case~II, where the sparse SN1987A data could not exclude the un-attenuated heavy-flavor scenario. 
The projection for the mass-proportional coupling is the most striking, because $\lambda_{\phi\nu}(m_1)$ is the smallest of the three eigenstate couplings while the depletion is driven by the much larger $\lambda_{\phi\nu}(m_{2,3})$, the bound on $\lambda_{\phi\nu}(m_1)$ reaches down to $\sim10^{-9}$. A future galactic supernova would thus probe couplings far beyond current reach and close the loophole that leaves the oscillated limit weak today.

\section{Conclusions}\label{sec:conclusions}

In this work we investigated the attenuation of supernova neutrinos by resonant scattering with the cosmic neutrino background through the process $\nu+\bar{\nu}\rightarrow\phi\rightarrow\chi+\bar{\chi}$. We derived the attenuation rate for both relativistic and non-relativistic relic neutrinos, incorporated it into the SN-neutrino flux reaching Earth, and included the in-medium flavor conversion that shapes the detected electron-antineutrino signal. The scalar coupling to neutrinos is diagonal in the mass basis and, depending on its ultraviolet origin, is assumed to be either mass-independent as in an inverse seesaw, or proportional to the neutrino mass as in a Type-I seesaw.

Fitting the SN1987A data with two emission models, the parameterized Pagliaroli09 and the hydrodynamic Vartanyan23, we found that the constraint on $\lambda_{\phi\nu}$ depends on both the flavor treatment and the coupling structure. In the flavor-blind case the two emission models give matching bounds. Once flavor conversion is included, the case with a mass-independent coupling exhibits a strong model dependence. In Vartanyan23, the sizable early-time heavy-flavor flux remains essentially unattenuated and feeds into the detected $\bar{\nu}_e$ channel, masking the depletion and consequently weakening the constraint from the sparse data. By contrast, Pagliaroli09 predicts only a small early-time heavy-flavor flux and is therefore largely unaffected by this masking effect. For the case with a mass-proportional coupling, the heavier mass eigenstates are strongly depleted instead. The masking effect disappears, and the two models again yield comparable constraints.
We further project the sensitivity of Hyper-Kamiokande, for which ${\sim}2000$ events from a Galactic supernova located at the distance of SN 1987A would improve the reach by one to two orders of magnitude and, crucially, restore a meaningful bound  even in regions where the current limit with flavor oscillations loses sensitivity.

Our results demonstrate that robust constraints on such neutrino interactions require modeling both the flavor conversion and the heavy-flavor emission, and that the resulting sensitivity depends on the ultraviolet origin of the coupling. 
Neglecting oscillations, or assuming a single coupling structure, can give a misleading estimate of the sensitivity, particularly for a sparse data set such as SN1987A. A future high-statistics supernova observation, together with improved modeling of the flavor-dependent emission, will therefore be crucial for probing resonant neutrino interactions with the C$\nu$B.

\section*{acknowledgments}
We thank Peizhi Du, Sida Lu, David Vartanyan, Tianshu Wang  and Yiming Zhong for helpful discussions on the project.
C.G.'s work is supported by the NSFC (Grant No. K25201009) and GDST (Grant No. K25203301).

\appendix

\section{Seesaw constructions}\label{ap:seesaw}

The effective coupling in Eq.~\eqref{lagrangian} can descend from different ultraviolet completions of neutrino mass~\cite{Minkowski:1977sc,Mohapatra:1979ia,Yanagida:1979as,Gell-Mann:1979vob,Schechter:1980gr,Mohapatra:1986bd,Ma:1987zm,Ma:2009gu,Bazzocchi:2010dt,Dias:2012xp,Law:2013gma}. Here we detail the two representative realizations, the minimal Type-I seesaw and the inverse seesaw.

\subsection{Type-I seesaw}

One may introduce three generations of right-handed sterile neutrinos, $N_{R,i}$, with Majorana masses~\cite{deGouvea:2016qpx}. The relevant Lagrangian is
\begin{equation}
   \mathcal{L} \supset y^{\alpha i} L_\alpha H N^c_{R,i} - \dfrac{1}{2} M_N^{ij} N_{R,i}^c N_{R,j}^c +  {\rm h.c.} \ .
\end{equation}
Here $\alpha = e,\mu,\tau$ denotes the lepton flavor and $i = 1,2,3$ stands for the sterile neutrino generations. The first term is the Yukawa coupling to the Higgs boson and the second term contains the Majorana mass matrix which can be rotated to the diagonal matrix.  
After electroweak symmetry breaking, the non-vanishing Higgs vacuum expectation value (VEV) can generate the Dirac mass connecting the active neutrino $\bar{\nu}_L$ and the sterile neutrino $N_R$ with the mass matrix $m_D = y v/\sqrt{2}$. 
The full neutrino mass matrix can then be diagonalized by rotating from the interaction basis $(\nu_\alpha,N^c_{R,i})$ to the mass basis $(\nu_l,\nu_h)$, where $\nu_l$ and $\nu_h$ denote the light and heavy neutrino mass eigenstates, respectively.
In the seesaw limit, $m_D\ll M_N$, the light-neutrino mass matrix is approximately given by $m_\nu \simeq m_D^T M_N^{-1} m_D$. 
The interaction eigenstates can be expressed in terms of the mass eigenstates as
%and the mixing can be expressed as
%
\begin{align}
\nu_{\alpha} &\approx (U_{\rm PMNS})_{\alpha j} \left( \nu_l + i \sqrt{\boldsymbol{M}_N^{-1}} \sqrt{\boldsymbol{m}_{\nu}} \, \nu_h \right)_j , \nonumber \\
N_{R,j}^c &\approx \left( \nu_h - i \sqrt{\boldsymbol{M}_N^{-1}} \sqrt{\boldsymbol{m}_{\nu}} \, \nu_l \right)_j \ ,
\end{align}
where we set the complex orthogonal Casas-Ibarra matrix $R =I$ for simplicity.  Restricting to a single neutrino flavor, the active-sterile mixing angle is approximately as $ \Theta \simeq (m_\nu/M_N)^{1/2}$. 

To get an expression for $\lambda_{\phi \nu}$ in Eq.~\eqref{lagrangian}, we introduce the following interaction,
\begin{equation}
\mathcal{L} \supset -\frac12\lambda_{\phi N} \phi N^c_{R,i}N^c_{R,i}+ h.c.
\end{equation}
This gives
\begin{equation}
\lambda_{\phi \nu}=\lambda_{\phi N} m_\nu/M_N \ .
\end{equation}
Note that $\lambda_{\phi \nu}$ has a structure proportional to the neutrino mass and is diagonal in the mass basis of active neutrinos.

%%%%%%%%%%%%%%%%%%%%%%%%%%%%%%%%%%
\subsection{Inverse seesaw}

In the inverse-seesaw scenario, two additional sterile fermions, $N^c_{R,i}$ and $S_{L,i}$, are introduced for each generation. 
%Compared with the conventional Type-I seesaw, the inverse seesaw allows the new fermions to remain close to the electroweak or TeV scale while simultaneously accommodating the observed smallness of the active-neutrino masses. This feature makes the inverse seesaw particularly attractive from a phenomenological perspective, since the sterile states may have sizable mixings with the active neutrinos and can potentially be probed in laboratory experiments.
The relevant Lagrangian is given by
\begin{equation}
    \mathcal{L}\supset
-y_{\alpha i}\, L_\alpha\,H\,N^c_{Ri}
-
\bar{S}_{Li}\,M_{ij}\,N_{Rj}
-
\frac12\,\mu^{ij} S_{L,i} S_{L,j}
+\text{h.c.},
\end{equation}
where $\alpha=(e,\mu,\tau)$ labels the SM lepton flavors, $i,j=1,2,3$ denote the sterile-fermion generations. 
The first term describes the Yukawa interaction among the SM lepton doublets, the Higgs field, and the right-handed neutrinos $N_R$.
The second term introduces a Dirac mass matrix $M$ connecting $S_L$ and $N_R$,
whereas the last term corresponds to a Majorana mass matrix $\mu$ for $S_L$.
After electroweak symmetry breaking, the Higgs vacuum expectation value generates the Dirac mass matrix $m_D = y v/2$ which mixes the active neutrinos with $N_R$. 
In the basis
$n_L=(\nu_L,\; N_R^c,\; S_L)^T$, the neutrino mass term can be written as
\begin{equation}
\mathcal{L}_{\rm mass}
=
-\frac{1}{2}\,\overline{n_L^c}\,
\mathcal{M}_\nu\,n_L+\text{h.c.},
\end{equation}
where the full $9\times9$ neutrino mass matrix takes the block form as 
\begin{equation}
\mathcal{M}_\nu=
\begin{pmatrix}
0 & m_D & 0\\
m_D^T & 0 & M^T\\
0 & M & \mu
\end{pmatrix}.
\end{equation}
Here $m_D$, $M$, and $\mu$ are general $3\times3$ matrices in flavor space.  
For $|\mu|\ll |m_D| \ll |M|$, the mass matrix diagonalization gives
\begin{equation}
m_\nu
\simeq
m_D (M^T)^{-1}\,\mu\,M^{-1}m_D^T ,
\end{equation}
up to higher-order corrections in $m_D/M$. 
Assuming that the heavy-sector matrices are approximately diagonal,
%
% \begin{equation}
% m_D={\rm diag}(m_{D_i}),
% \qquad
% M={\rm diag}(M_i),
% \qquad
% \mu={\rm diag}(\mu_i),
% \end{equation}
% %
the light neutrino masses are approximately
\begin{equation}
m_{\nu_i}
\simeq
\frac{m_{D_i}^2}{M_i^2}\,\mu_i .
\end{equation}
The smallness of the active-neutrino masses is therefore primarily controlled by the small lepton-number-violating parameter $\mu$, rather than solely by the heaviness of the sterile states. Thus the light-neutrino masses vanish in the technically natural limit $\mu\to0$, where lepton number is restored.
%In contrast to the Type-I seesaw, achieving the light-neutrino masses does not require the sterile-neutrino mass scale $M$ to be extremely large.

The full $9\times9$ mass matrix is diagonalized by a unitary matrix $U$,
\begin{equation}
 U^T\mathcal{M}_\nu U =
 \mathrm{diag}(m_1,m_2,m_3,M_1,\ldots,M_6).
 \end{equation}
$N_R$ and $S_L$ combine into three approximately Dirac fermions with masses close to $M$. 
A small $\mu$ splits each Dirac fermion into a pair of nearly degenerate Majorana states, often referred to as a pseudo-Dirac fermion pair. 
In the limit $\mu_i\to0$ and $m_{D_i}\ll M_i$, the mass eigenstates are
\begin{align}
\nu_{l, i} &\simeq \nu^{(0)}_{L,i}- \frac{m_{D_i}}{M_i} S_{L,i}~,
\\
\nu_{h,i}^+
&\simeq
\frac{1}{\sqrt2}
\left(
\frac{m_{D_i}}{M_i}\nu^{(0)}_{L,i}
+
N_{R,i}^c
+
S_{L,i}
\right),\\
\nu_{h,i}^-
&\simeq
\frac{1}{\sqrt2}
\left(
\frac{m_{D_i}}{M_i}\nu^{(0)}_{L,i}
-
N_{R,i}^c
+
S_{L,i}
\right)~,
\end{align}
where $\nu^{(0)}_{L}=U_{\rm PMNS}^\dagger \nu_L$. 
Inverting these relations, one obtains
\begin{align}
\nu_{L,\alpha} &
\simeq
\sum_{i=1}^3
(U_{\rm PMNS})_{\alpha i}
\left[
\nu_{l,i}
+
\frac{m_{D_i}}{M_i}
\frac{\nu_{h,i}^++\nu_{h,i}^-}{\sqrt2}
\right],\\
N_{R,i}^c &
\simeq
\frac{1}{\sqrt2}
\left(
\nu_{h,i}^+ - \nu_{h,i}^-
\right),\\
S_{L,i} &
\simeq
-
\frac{m_{D_i}}{M_i} \nu_{l,i}
+
\frac{1}{\sqrt2}
\left(
\nu_{h,i}^+ + \nu_{h,i}^-
\right).
\end{align}
The mixing angle between the sterile neutrino and active neutrino is around $m_{D_i}/M_i$.

To recover $\lambda_{\phi \nu}$ in \eqref{lagrangian}, one can introduce the following interaction:
\begin{equation}
\mathcal{L} \supset -\frac12\,\lambda_{\phi S} \,\phi S_{L,i}S_{L,i}+ {\rm h.c.} \ .
\end{equation}
This gives
\begin{equation}
\lambda_{\phi \nu}=\lambda_{\phi S} (m_{D_i}/M_i)^2
\end{equation}
Unlike Type-I seesaw, the value of $\lambda_{\phi \nu}$ is generally independent of the neutrino mass.  However, it can still be made diagonal in the mass basis of active neutrinos.

\section{Cross section and attenuation rate}\label{ap:xsec}

For the scattering process $\nu(p)\bar\nu(k)\to\phi^\ast\to\chi(p') \chi(k')$, the tree amplitude from Eq.~\eqref{lagrangian} is
\begin{equation}
\mathcal{M}=-i\,\lambda_{\phi\nu}\lambda_{\phi\chi}\,
\frac{[\bar v(k)u(p)]\,[\bar u(p')v(k')]}{s-m_\phi^2+im_\phi\Gamma_\phi},
\end{equation}
keeping the mass $m_\nu$ on both incoming legs and $m_\chi=0$. Averaging over the initial helicities and using the scalar-coupling traces $\mathrm{Tr}[(\slashed{p}+m_\nu)(\slashed{k}-m_\nu)]=2(s-4m_\nu^2)$ and $\mathrm{Tr}[\slashed{p}'\slashed{k}']=2s$, we get
\begin{equation}\label{eq:M2}
\overline{|\mathcal{M}|^2}=\lambda_{\phi\nu}^2\lambda_{\phi\chi}^2\,\frac{s(s-4m_\nu^2)}{(s-m_\phi^2)^2+m_\phi^2\Gamma_\phi^2}.
\end{equation}
The factor $(s-4m_\nu^2)$ is the $p$-wave threshold characteristic of a scalar (as opposed to pseudoscalar) coupling.

With the invariant flux $F=4\sqrt{(p\cdot k)^2-m_\nu^4}=2\sqrt{s(s-4m_\nu^2)}$ and the isotropic two-body phase space $\int d\Pi_2=1/8\pi$, one recovers Eq.~\eqref{eq:sigma},
\begin{equation}
\sigma=\frac{1}{F}\int\overline{|\mathcal{M}|^2}\,d\Pi_2
=\frac{\lambda_{\phi\nu}^2\lambda_{\phi\chi}^2}{16\pi}\,\frac{\sqrt{s(s-4m_\nu^2)}}{(s-m_\phi^2)^2+m_\phi^2\Gamma_\phi^2}.
\end{equation}
The on-shell width of $\phi\to\chi\chi$ (massless, Majorana, with the $1/2$ symmetry factor for identical final states) is $\Gamma_\phi=\lambda_{\phi\chi}^2m_\phi/16\pi$. This normalization is consistent with the width $\delta E_\nu$ in Eq.~\eqref{eq:dEk}.

The M\o ller velocity is $v_{\rm M\o l}=\sqrt{(p\cdot k)^2-m_\nu^4}/(E_1E_\nu)$, so the flux cancels in the rate kernel,
\begin{equation}
\sigma\,v_{\rm M\o l}=\frac{1}{4E_1E_\nu}\int\overline{|\mathcal{M}|^2}\,d\Pi_2
=\frac{\lambda_{\phi\nu}^2\lambda_{\phi\chi}^2}{32\pi}\,\frac{s(s-4m_\nu^2)}{E_1E_\nu\,[(s-m_\phi^2)^2+m_\phi^2\Gamma_\phi^2]}.
\end{equation}
In the SN-relativistic limit ($E_\nu\gg m_\nu$, $s\simeq2E_1E_\nu(1-\beta\cos\theta)$, $s-4m_\nu^2\simeq s$) this reduces to Eq.~\eqref{eq:sigmav}.

%Relic neutrinos decouple while relativistic, so $f_{\rm FD}$ is a function of the momentum $|\vec p\,|$. Using $d^3p/(2\pi)^3=|\vec p\,|^2 d|\vec p\,|\,d\cos\theta/(4\pi^2)$ with the energy variable $x=E_1/T_\nu$ (so $|\vec p\,|=T_\nu\sqrt{x^2-x_0^2}$), Eq.~\eqref{eq:rate} yields Eqs.~\eqref{eq:scattering_rate}--\eqref{eq:normalized_rate}. The angular weight is $(1-\beta\cos\theta)^2$, inherited from $\sigma\,v_{\rm M\o l}$.

\section{SN neutrino flux}\label{ap:flux}

During core-collapse supernova (CCSN) supernova explosion, neutrinos carry away about $99\%$ of the emitted energy. However, accurate modeling of this process remains challenging due to immense complexity of the underline physics. 

In this work, we primarily consider the more detailed neutrino-driven mechanism~\cite{Bethe:1985sox, Loredo:2001rx, pagliaroli2009improved}. 
Within this mechanism, three processes contribute to the total neutrino emission. 
The late-time cooling phase accounts for approximately $80\sim 90\%$ of the emitted neutrino energy, dominating the neutrino budget. Preceding the cooling phase, the accretion phase exhibits a much higher luminosity but lasts shorter,
contributing to $10\sim20\%$ of the total energy. Moreover, an initial neutronization burst occurs prior to the two main phases. Since it releases only $\sim 1\%$ of the total energy and has a negligible effect on the expected signal rate, we choose to ignore this from our analysis. 

To comprehensively evaluate neutrino flux, we compare two emission models. 
The first one is the parameterized Pagliaroli09~\cite{pagliaroli2009improved}, the second one is the-state-of-art, 2D hydrodynamic Vartanyan23~\cite{Vartanyan:2023zlb}. In this section, we analyze each flux component and discuss both models in more details.

%%%%%%%%%%%%%%%%%
\subsection{Pagliaroli09}\label{ap:pmodel}

The emission is described by two successive stages~\cite{pagliaroli2009improved}. The \emph{cooling phase} carries most of the energy and is quasi-thermal. The differential $\bar\nu_e$ flux on Earth, from a source at distance $d_{\rm SN}$, is
\begin{equation}\label{eq:cooling_flux}
\frac{d\Phi_{c, \,\bar\nu_e}^0}{dE_\nu}(t, E_\nu) =\frac{R_c^2}{d_{\rm SN}^2}\frac{\pi c}{(2\pi)^3}\,f_\nu(t,E_\nu)\, E_\nu^2 ,
\end{equation}
with a Fermi--Dirac spectrum $f_\nu=[1+\exp(E_\nu/T_c(t))]^{-1}$ whose temperature decreases as $T_c(t)=T_c\,e^{-t/4\tau_c}$. This stage is controlled by the neutrinosphere radius $R_c$ (comparable to the remnant neutron-star radius), the initial temperature $T_c\sim3$--$6\,$MeV, and the timescale $\tau_c\sim\mathcal{O}(1)\,$s~\cite{pagliaroli2009improved}.

The shorter but more luminous \emph{accretion phase} occurs while matter accretes onto the stalled shock, and its $\bar\nu_e$ emission is dominated by positron capture on free neutrons, $e^++n\to p+\bar\nu_e$. 
%Following Ref.~\cite{pagliaroli2009improved}, 
Its flux $\Phi_{a, \,\bar\nu_e}^0$ is parameterized by the accreting mass $M_a$, the initial temperature $T_a$, and the accretion timescale $\tau_a\sim0.5\,$s, with the positron-capture cross section, quasi-thermal positron distribution, and effective free-neutron number taken from that reference.

The two stages are combined with a time-shift~\cite{pagliaroli2009improved},
\begin{equation}\label{eq:flux_nu_e}
\Phi_{\bar\nu_e}^0(t) = j_k(t)\,\Phi_{a, \,\bar\nu_e}^0(t) + [1 - j_k(t)]\,\Phi_{c, \,\bar\nu_e}^0(t-\tau_a),
\qquad j_k(t)=e^{-(t/\tau_a)^2},
\end{equation}
so that the accretion component dominates for $t\lesssim\tau_a$ and the cooling component for $t\gtrsim\tau_a$. The continuity across the transition is maintained by extending $\Phi_{c,\bar\nu_e}^0$ to slightly negative time arguments. In total, Pagliaroli09 carries six source parameters, $\{R_c,T_c,\tau_c,M_a,T_a,\tau_a\}$.

During the cooling phase, the emitted energy is assumed to be equally partitioned among flavors, while heavy-flavor antineutrinos decouple deeper within the proto-neutron star and therefore have a higher temperature. Following Ref.~\cite{keil2003monte}, we adopt $T_c(\bar{\nu}_x)/T_c(\bar{\nu}_e)=1.2$, assuming a common neutrinosphere radius. During the accretion phase, by contrast, the emission is dominated by electron neutrinos and antineutrinos through the charged-current capture processes $e^-+p\to n+\nu_e$ and $e^++n\to p+\bar{\nu}_e$, respectively.
%; introducing an independent heavy-flavor component there would require extra source parameters and induce degeneracies given the sparse SN1987A data.
Hence, we have
\begin{equation}\label{eq:flux_nu_x}
\Phi_{{\bar\nu}_x}^0(t) = [1 - j_k(t)]\,\Phi_{c,\,{\bar\nu}_x}^0(t-\tau_a),
\end{equation}
with $\Phi_{c,{\bar\nu}_x}^0$ the cooling-phase flux in Eq.~\eqref{eq:cooling_flux} evaluated at the heavy-flavor temperature.

\subsection{Vartanyan23}\label{ap:vmodel}

For comparison we use the 2D radiation-hydrodynamic simulations of Ref.~\cite{Vartanyan:2023zlb}, a suite of 100 core-collapse runs spanning a range of progenitor masses. Unlike Pagliaroli09, which fits the SN1987A data through a few interpretable quantities, these employ the \texttt{FORNAX} code to evolve the neutrino transport in discrete energy groups, so the time- and energy-dependent spectra emerge from first principles rather than being assumed Fermi--Dirac. In particular, the simulations consistently produce heavy-flavor neutrinos through the neutral-current pair processes $e^-+e^+\to\nu_x+\bar\nu_x$ and $N+N\to N+N+\nu_x+\bar\nu_x$, yielding a sizable $\bar\nu_x$ flux already during the early accretion stage.

Although the runs span many progenitor masses, the neutrino signals correlate most directly with the compactness parameter
\begin{equation}
\xi_M=\frac{M/M_\odot}{R(M)/1000\,{\rm km}},
\end{equation}
with $R(M)$ the radius enclosing baryonic mass $M$. The luminosities, mean energies, and accretion rates depend more monotonically on $\xi_M$ than on the progenitor mass itself. Since SN1987A presumably originated from a $15$--$20\,M_\odot$ progenitor, we use the simulated spectra in this range, treating their variation as a sampling of progenitor structure (compactness) rather than a mass uncertainty.

The tabulated spectra are interpolated in time and energy to give continuous fluxes. For the electron antineutrino,
\begin{equation}\label{eq:flux_nu_e_v}
    \Phi_{\bar\nu_e}^0(t) = A\int_0^{E_{\rm max}} \frac{1}{4\pi d_{\rm SN}^2}\frac{d\Phi_{\bar\nu_e}}{dE}\,dE,
\end{equation}
where $A$ is a unit-conversion factor and $E_{\rm max}=50$ MeV sets the bounds on the interpolation range. The heavy-flavor species are simulated as a single group, so the per-species flux is obtained by dividing by four,
\begin{equation}\label{eq:flux_nu_x_v}
    \Phi_{\bar\nu_x}^0(t) = \frac{A}{4}\int_0^{E_{\rm max}}\frac{1}{4\pi d_{\rm SN}^2}\frac{d\Phi_H}{dE}\,dE.
\end{equation}
The simulations cover only the first $4$--$5\,$s post-bounce (the pre-bounce flux at $t<0$ is negligible, so $t=0$ marks the onset, aligned with the first detected event). We therefore discard events after $4\,$s, leaving $17$ events for this model, compared with $24$ for Pagliaroli09.

\subsection{Matter effect}\label{ap:flavor}

Here we derive the mapping between the emitted flavor states and the propagating vacuum mass eigenstates used in Sec.~\ref{sec:n_osc}, following the standard MSW treatment of SN neutrinos~\cite{Dighe:1999bi,mirizzi2016supernova} specialized to the antineutrino channel in the normal mass ordering.

In the flavor basis $(\bar\nu_e,\bar\nu_\mu,\bar\nu_\tau)$ the antineutrino propagation Hamiltonian is
\begin{equation}
\bar H=\frac{1}{2E_\nu}\,U^{*}\,{\rm diag}(m_1^2,m_2^2,m_3^2)\,U^{T}-{\rm diag}(V_e,0,0),
\end{equation}
where $U$ is the PMNS lepton mixing matrix ($\nu_\alpha=\sum_i U_{\alpha i}\nu_i$), $V_e=\sqrt2\,G_F n_e$ is the charged-current matter potential, and, crucially, the matter term carries a minus sign for antineutrinos (opposite to the neutrino case).

Deep in the core the electron density is is sufficiently high that $2E_\nu V_e\gg|\Delta m^2_{ij}|$~\cite{Dighe:1999bi,mirizzi2016supernova}. The matter term $-V_e$ therefore dominates, and $\bar H$ is approximately diagonal in the flavor basis. The instantaneous (matter) eigenstates therefore coincide with the flavor states to leading order. 
Because the dominant entry is negative, the $\bar\nu_e$-like state carries the extreme eigenvalue $\simeq-V_e$ and is the lowest matter eigenstate, while $\bar\nu_\mu,\bar\nu_\tau$ lie near zero. We label each matter eigenstate $\bar\nu_{im}$ by the vacuum mass eigenstate $\bar\nu_i$ to which it evolves as $n_e\to0$. For normal mass ordering, no MSW resonance occurs in the antineutrino channel, so the level ordering is preserved as the density decreases. The lowest matter eigenstate therefore connects continuously to the lightest vacuum mass eigenstate, $\bar\nu_1$. Thus, in the high-density limit, $\bar\nu_e=\bar\nu_{1m}$ and $\bar\nu_{\mu,\tau}=\bar\nu_{2m,3m}$.

The evolution outward is adiabatic when the density scale height $|d\ln n_e/dr|^{-1}$ is much larger than the local oscillation length. 
This condition is satisfied in the SN envelope for the energies considered here. In the absence of an MSW resonance, antineutrinos therefore remain on their respective instantaneous matter-eigenstate branches as the density decreases~\cite{Dighe:1999bi}.
Each matter eigenstate thus emerges as the corresponding vacuum eigenstate, $\bar\nu_{im}\to\bar\nu_i$. In particular, an electron antineutrino produced as $\bar\nu_{1m}$ exits the star as $\bar\nu_1$, consistent with the identification in Eq.~\eqref{eq:matter_id}. 

Over the propagation distance $d_{\rm SN}\simeq50\,$kpc, the mass-eigenstate wave packets separate well beyond their coherence length and therefore arrive at Earth as an incoherent mixture. In inverse beta decay the $\bar\nu_1$ component interacts as $\bar\nu_e$ with probability $|U_{e1}|^2$, giving the survival probability $\bar p=|U_{e1}|^2$ and hence the detected flux of Eq.~\eqref{eq:mixing_flux_detector}. This assignment is specific to the antineutrino channel in the normal ordering. The resonance structure and resulting flavor mapping differ for neutrinos and for antineutrinos in the inverted ordering.

\section{Detection and likelihood analysis}\label{ap:detection}

SN1987A occurred in the Large Magellanic Cloud, at a distance of $51.4\pm1.2$ kpc from light-curve and SN-ring measurements~\cite{panagia1999distance, panagia2005geometric} and $49.59\pm0.09_{\rm stat}\pm0.54_{\rm syst}$ kpc from a more recent geometric determination~\cite{pietrzynski2019distance}. We adopt $d_{\rm SN}=50$ kpc.

Four independent experiments reported the detection of neutrinos at the time of the SN1987A explosion. Among these, Kamiokande-II (Kam-II) recorded the largest dataset, consisting of 16 overall events~\cite{Kamiokande-II:1987idp, Kamiokande-II:1989hkh}. However, the sixth and the last four events (K6, K13–K16) are widely considered to be background contamination and are excluded from our analyses. Meanwhile, the Irvine-Michigan-Brookhaven (IMB)~\cite{bionta1987observation, bratton1988angular} and BUST (Baksan)~\cite{Alekseev:1988gp} detectors reported 8 and 5 events, respectively. The Liquid Scintillator Detector (LSD) at Mont Blanc also reported 5 events~\cite{dadykin1987detection}. However, this detection occurred approximately 4.7 hours prior to the others. Because these early events cannot be associated with the signals seen by IMB, Kamiokande-II, and Baksan within the context of standard core-collapse scenarios, they are excluded from our joint analysis. All the events used in this work, along with their measured properties, are listed in Table \ref{tab:events}.

\begin{table}[]
    \centering
    \begin{tabular}{ccccc}
    \hline
    \hline
         Event No.& Time   & Energy         & Angle         & Background          \\
                  &[sec]   &[MeV]           &[Degree]       &[$\rm MeV^{-1}s^{-1}$]     \\
         \hline

         K1       &$\equiv 0$ & $20   \pm 2.9$ & $18  \pm 18$  & $1\times10^{-5}$    \\
         K2       & 0.107  & $13.5 \pm 3.2$ & $40  \pm 27$  & $5.4\times10^{-4}$  \\
         K3       & 0.303  & $7.5  \pm 2.0$ & $108 \pm 32$  & $3.1\times10^{-2}$  \\
         K4       & 0.324  & $9.2  \pm 2.7$ & $70  \pm 30$  & $8.5\times10^{-3}$  \\
         K5       & 0.507  & $12.8 \pm 2.9$ & $135 \pm 23$  & $5.3\times10^{-4}$  \\
         K7       & 1.541  & $35.4 \pm 8.0$ & $32  \pm 16$  & $5\times10^{-6}$    \\
         K8       & 1.728  & $21.0 \pm 4.2$ & $30  \pm 18$  & $1\times10^{-5}$    \\
         K9       & 1.915  & $19.8 \pm 3.2$ & $38  \pm 22$  & $1\times10^{-5}$    \\
         K10      & 9.219  & $8.6  \pm 2.7$ & $122 \pm 30$  & $1.8\times10^{-2}$  \\
         K11      & 10.433 & $13.0 \pm 2.6$ & $49  \pm 26$  & $4\times10^{-4}$    \\
         K12      & 12.439 & $8.9  \pm 1.9$ & $91  \pm 39$  & $1.4\times10^{-2}$  \\
         
         \hline
         I1       &$\equiv 0$ & $38   \pm 7$   & $80  \pm 10$  & 0                   \\
         I2       & 0.412  & $37   \pm 7$   & $44  \pm 15$  & 0                   \\
         I3       & 0.650  & $28   \pm 6$   & $56  \pm 20$  & 0                   \\
         I4       & 1.141  & $39   \pm 7$   & $65  \pm 20$  & 0                   \\
         I5       & 1.562  & $36   \pm 9$   & $33  \pm 15$  & 0                   \\
         I6       & 2.684  & $36   \pm 6$   & $52  \pm 10$  & 0                   \\
         I7       & 5.010  & $19   \pm 5$   & $42  \pm 20$  & 0                   \\
         I8       & 5.582  & $22   \pm 5$   & $104 \pm 20$  & 0                   \\
         
         \hline
         B1       &$\equiv 0$ & $12.0 \pm 2.4$ & $\equiv 90$   & $8.4\times10^{-4}$  \\
         B2       & 0.435  & $18.0 \pm 3.6$ & $\equiv 90$   & $1.3\times10^{-3}$  \\
         B3       & 1.710  & $23.3 \pm 4.7$ & $\equiv 90$   & $1.2\times10^{-3}$  \\
         B4       & 7.687  & $17.0 \pm 3.4$ & $\equiv 90$   & $1.3\times10^{-3}$  \\
         B5       & 9.099  & $20.1 \pm 4.0$ & $\equiv 90$   & $1.3\times10^{-3}$  \\
         \hline
    \end{tabular}
    \caption{SN1987A events used in our analysis. Column 1: event number and detector notion, with K the Kamiokande-II, I the IMB and B the Baksan; Column 2: time sequence relative to the first detection in each experiment; Column 3-4: energies and angles with their errors; Column 5: Estimated number density due to background.}
    \label{tab:events}
\end{table}

The IBD differential cross section is computed following Ref.~\cite{strumia2003precise},
\begin{equation}
\frac{d\sigma}{dt}=\frac{G_F^2\cos^2\theta_C}{2\pi(s-m^2_p)^2}|\mathcal{M}|^2,
\end{equation}
with $s=(p_\nu+p_p)^2$, $t=(p_\nu-p_e)^2$, and $|\mathcal{M}|^2$ a function of the Mandelstam invariants and the nucleon form factors. In the proton rest frame,
\begin{equation}\label{eq:t}
t=m_n^2-m_p^2-2m_p(E_\nu-E_e)=m_e^2-2E_\nu(E_e-p_e\cos\theta),
\end{equation}
which leads to the cross section in the measured variables,
\begin{equation}
\frac{d\sigma}{d\cos\theta}= \frac{d\sigma}{dt}\,\frac{2E_\nu p_e}{1-\varepsilon\left(\frac{E_e}{p_e}\cos\theta-1\right)},\qquad \varepsilon\equiv E_\nu/m_p \ .
\end{equation}
Fixing the incident energy,
\begin{equation}
E_\nu=\frac{\delta+E_e}{1-\left(E_e-p_e\cos\theta\right)/m_p},\qquad \delta\equiv \frac{m_n^2-m_p^2-m_e^2}{2m_p}.
\end{equation}
%or
%\begin{equation}
%E_e=\frac{( E_\nu -\delta ) (\varepsilon +1)+\varepsilon c_\theta  \sqrt{(\delta -E_\nu)^2-m_e^2 \left(\varepsilon ^2 \sin^2\theta+2 \varepsilon +1\right)}}{\varepsilon ^2 \sin^2 \theta+2 \varepsilon +1}
%\end{equation}

Each detector is characterized by its effective number of target protons $N_p$ (set by the fiducial mass), the angular bias $\xi(c_\theta)$, the energy threshold $E_{\rm th}$, and the energy-dependent efficiency $\eta(E_e)$, as summarized in Table~\ref{tab:detectors}~\cite{Kamiokande-II:1987idp, Kamiokande-II:1989hkh, bionta1987observation,bratton1988angular, Alekseev:1988gp}. The efficiency does not reach $100\%$ immediately above threshold, owing to detector geometry, photoelectron statistics, and radioactivity~\cite{Loredo:2001rx, Costantini:2004ry}. We adopt the analytic Kam-II and IMB fits of Ref.~\cite{Burrows_1988} and take a constant $0.8$ for Baksan,
\begin{widetext}
\begin{equation}\label{eq:efficiency}
\begin{split}
    \eta_{\rm KAM-II}(E_e) &= 0.93 - e^{-\left(\frac{E_e}{ 9~\rm MeV} \right)^{2.5}}, \qquad E_e>7 ~ \rm MeV; \\
    \eta_{\rm IMB}(E_e) & = 0.3975\left(\frac{E_e}{ 10~\rm MeV} \right) - 0.02625\left(\frac{E_e}{ 10~\rm MeV} \right)^2 - 0.59, \qquad E_e>19 ~ \rm MeV; \\
    \eta_{\rm Baksan}(E_e) & = 0.8,\qquad  E_e>12 ~ \rm MeV.
\end{split}
\end{equation}
\end{widetext}

\begin{table}[htbp]
    \centering
    \begin{tabular}{lccc}
    \hline\hline
     & Kam-II & IMB & Baksan \\
    \hline
    $N_p$ & $1.43\times10^{32}$ & $4.55\times10^{32}$ & $1.87\times10^{31}$ \\
    $E_{\rm th}$ [MeV] & $7$ & $19$ & $12$ \\
    $\xi(c_\theta)$ & $1$ & $1+0.1\cos\theta$ & $1$ \\
    $\eta(E_e)$ & Eq.~\eqref{eq:efficiency} & Eq.~\eqref{eq:efficiency} & $0.8$ \\
    $f_d$ & $1$ & $0.9055$ & $1$ \\
    $\tau_d$ [s] & $0$ & $0.035$ & $0$ \\
    $N_{\rm ev}$ ($t<30\,$s\,/\,$t<4\,$s) & $11\,/\,8$ & $8\,/\,6$ & $5\,/\,3$ \\
    \hline\hline
    \end{tabular}
    \caption{Detector properties used in the analysis: the effective number of target protons $N_p$, energy threshold $E_{\rm th}$, angular bias $\xi(c_\theta)$, and efficiency $\eta(E_e)$ (Eq.~\eqref{eq:efficiency}), the live-time fraction $f_d$ and dead time $\tau_d$, together with the number of events $N_{\rm ev}$ entering the Pagliaroli09 ($t<30\,$s) and Vartanyan23 ($t<4\,$s) analyses.}
    \label{tab:detectors}
\end{table}

The event times in Table~\ref{tab:events} are measured relative to the first event in each detector and may differ from the absolute onset by an offset, $t_i = t^{\rm off}+ \delta t_i$. For Kam-II and Baksan, $t^{\rm off}=0$ lies well within the $1\sigma$ region of global fits~\cite{pagliaroli2009improved, Loredo:2001rx}. For IMB, $t^{\rm off}_{\rm IMB}=0$ and $t^{\rm off}_{\rm IMB}\sim\tau_a$ give comparable $\chi^2$~\cite{pagliaroli2009improved}, but the latter requires an implausibly long accretion phase ($\tau_a>1\,$s) and a large accreting mass ($M_a>1\,M_\odot$). We therefore set $t^{\rm off}=0$ for all three detectors.

In the likelihood of Eq.~\eqref{eq:likelihood}, the expected event number is
\begin{equation}
    N_{\rm exp} = \int_0^{t_{\rm max}} f_dS(t) dt,
\end{equation}
and the theoretical differential event rate of the $i$-th event is
\begin{equation}
    \mathcal{R}(t_i, E_i, c_{\theta, i}) = e^{S(t_i) \tau_d} \times  \left[ \frac{B_i}{2} + \int_0^{E_{\rm max}} \frac{d S}{dE_{e}dc_\theta}(t_i,E_e,c_{\theta, i}) G(E_e, E_i) \mathrm{d}E_e \right].
\end{equation}
% \begin{equation}\label{eq:likelihood_full}
% \begin{split}
% &\mathcal{L}_d = e^{-f_d \int R(t) \mathrm{d}t} \times \\&\prod_{i=1}^{N_d}  e^{R(t_i) \tau_d} \times  \left[ \frac{B_i}{2} + \int \frac{d R}{dE_{e}dc_\theta}(t,E_e,c_\theta) G(E_e, E_i) \mathrm{d}E_e \right] 
% \end{split}
% \end{equation}
$B_i$ are the background values and for Kamiokande-II and Baksan are taken from Ref.~\cite{Costantini:2006xd} and~\cite{Loredo:2001rx}, respectively. For IMB, the background contamination is negligible within the relevant time window, so we adopt $B_i = 0$. The term $S(t)$ is the integration of Eq.~\eqref{eq:signalrate} over $E_e$ and $c_\theta$. The integration range of $E_e$ is from 0 to $E_{\rm max} = 50 ~{\rm MeV}$ , and for $c_\theta$ it is from $-1$ to 1. The integration range of $t$ is from $0$ to $t_{\rm max}=30~\mathrm{s}$ for Pagliaroli09 in consistency with Ref.~\cite{pagliaroli2009improved}, while for Vartanyan23 $t_{\rm max}$ is defined by the maximum cut-off time of each mass-based simulation. $f_d$ and $\tau_d$ are the per-detector live-time fraction and dead time, listed in Table~\ref{tab:detectors}. $G(E_e, E_i)$ accounts for the detector-specific energy resolution, and is modeled as a Gaussian function where $\sigma_{E,i}$ is the estimated energy uncertainty for the $i$-th event:
\begin{equation}
G(E_e, E_i) = \frac{1}{\sqrt{2 \pi} \sigma_{E,i}} e ^{- \frac{(E_e-E_i)^2} {2 \sigma_{E,i} ^ 2}}.
\end{equation}
%Here we don't include the errors of angle $c_i$ and event time $t_i$ in consistency with Ref.~\cite{pagliaroli2009improved}.

Parameter estimation is performed based on the MCMC ensemble sampler implemented in the Python package \texttt{emcee} ~\cite{foreman2013emcee}. The code for this analysis as well as MCMC chains is publicly available on GitHub\footnote{\url{https://github.com/rzvincent/sn1987_neutrino_mcmc}}. 
The priors are chosen to be uniform in a large range, listed in the second column of Table.~\ref{tab:best_fit}. 
%Since the resonance is only prominent when log($m_\phi/{\rm eV}$) falls in $(1,\, 3)$, here we considered the sub-keV range with a slightly broader range.
We conducted the MCMC procedure on both neutrino flux emission models. For Pagliaroli09, we employ an ensemble of 32 walkers, advancing each for $50,000$ steps to ensure robust convergence across the parameter space. The first $30,000$ steps of each chain are discarded as burn-in. The remaining samples are subsequently thinned by a factor of 20 to mitigate autocorrelation. Thus, flattening the arrays yields a final posterior sample of size $(32000, 8)$, where $8$ denotes the dimension of our parameter space.

\section{Posterior distributions and forecasts}\label{ap:stats}

The best-fit parameters and their $1\sigma$ C.L with Pagliaroli09 are listed in Table.~\ref{tab:best_fit}. For the new physics parameters, all the three cases are consistent with zero-coupling within $1\sigma$ C.L and don't show a preferred constraint. The credible levels of 1D posterior distributions are acquired by calculating the HPD.

\begin{table}[htbp]
    \centering
    \begin{tabular}{ccccc}
    \hline
    \hline
        $\theta$  & Priors & Case I & Case II & Case III \\
    \hline
        $R_c$/ $\rm [km]$   & $U(1, 100)$   & $10.42_{-5.71}^{+9.00}$ & $10.99_{-5.78}^{+10.82}$ & $11.65_{-6.60}^{+10.61}$ \\
        $T_c$/ $\rm [MeV]$  & $U(1, 10)$    & $4.50_{-0.57}^{+1.03}$  & $4.21_{-0.63}^{+0.85}$ & $4.24_{-0.61}^{+0.95}$ \\
        $\tau_c$/ $\rm [s]$ & $U(1, 10)$    & $4.61_{-1.28}^{+1.72}$  & $4.49_{-1.28}^{+1.56}$ & $4.48_{-1.24}^{+1.88}$ \\
        $M_a$/ $[M_\odot]$  & $U(0.001, 2)$ & $0.07_{-0.06}^{+0.56}$  & $0.11_{-0.08}^{+0.75}$ & $0.10_{-0.08}^{+0.83}$ \\
        $T_a$/ $\rm [MeV]$  & $U(1, 10)$    & $2.02_{-0.28}^{+0.32}$  & $1.98_{-0.19}^{+0.40}$ & $2.02_{-0.23}^{+0.35}$ \\
        $\tau_a$/ $\rm [s]$ & $U(0.1, 2)$   & $0.45_{-0.14}^{+0.64}$  & $0.45_{-0.18}^{+0.50}$ & $0.46_{-0.17}^{+0.70}$ \\
        log$(m_\phi)$/ $\rm [eV]$ & $U(-1, 4)$ & N/A &  N/A &  N/A \\
        log$(\lambda_{\phi\nu})$ & $U(-16, 0)$ &  N/A &  N/A &  N/A \\
    \hline
    \end{tabular}
    \caption{The best-fit parameters and their $1\sigma$ C.L with Pagliaroli09. The three cases are in the same manner as Fig.~\ref{fig:corner_plots}.}
    \label{tab:best_fit}
\end{table}

\begin{figure*}
    \centering
    \includegraphics[width=0.8\linewidth]{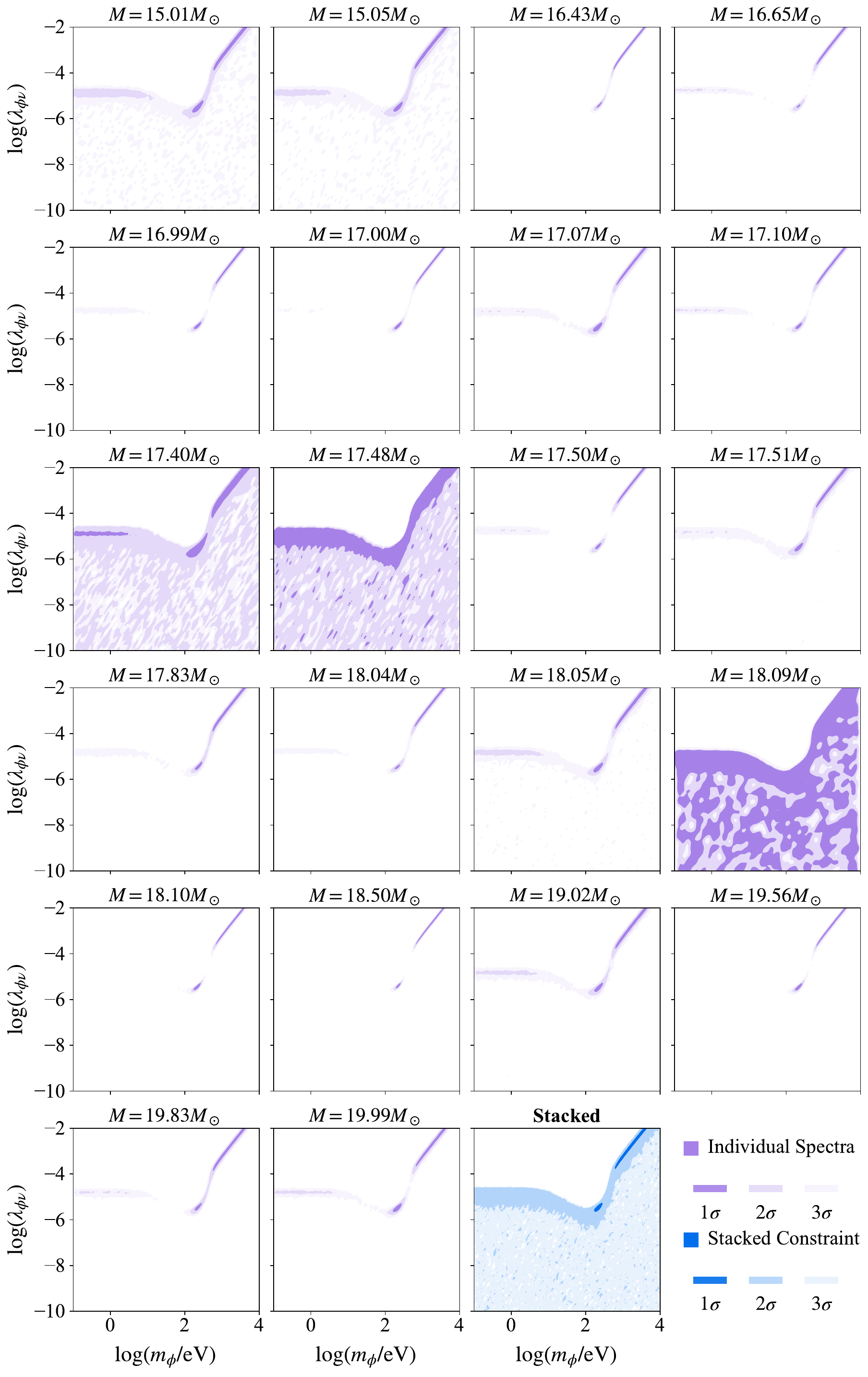}
    \caption{Vartanyan23: 2D contour plots of constraining 2 new physics parameters with 22 independent spectra based on Case I. The blue contour plot is derived by stacking the 22 constraints with the same weight. The shaded regions represent $1\sigma$(dark), $2\sigma$(medium), $3\sigma$(light) C.L. }
    \label{fig:2d_sim_all_I}
\end{figure*}

\begin{figure*}
    \centering
    \includegraphics[width=0.8\linewidth]{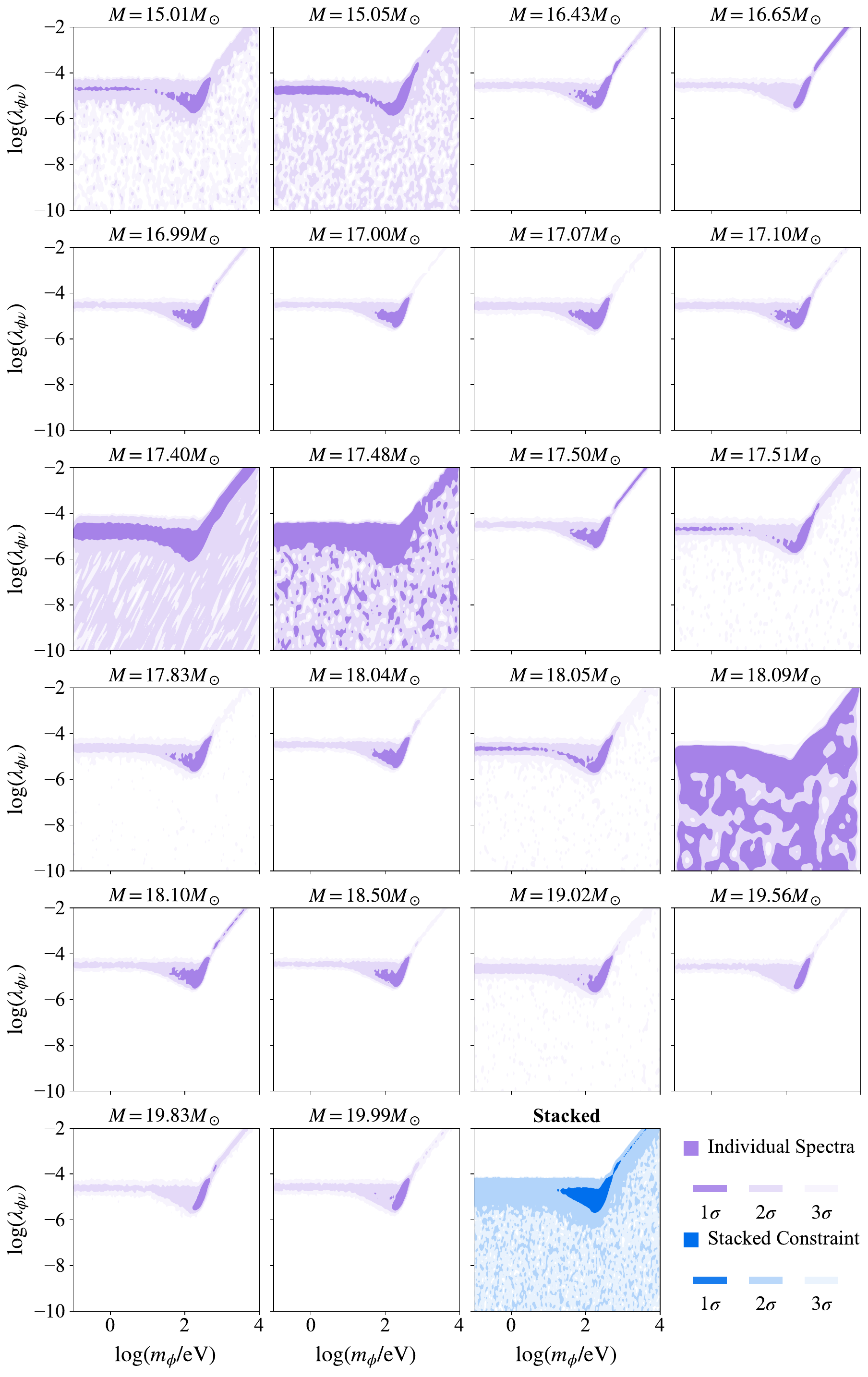}
    \caption{Vartanyan23: 2D contour plots of constraining 2 new physics parameters with 22 independent spectra based on Case II. }
    \label{fig:2d_sim_all_II}
\end{figure*}

\begin{figure*}
    \centering
    \includegraphics[width=0.8\linewidth]{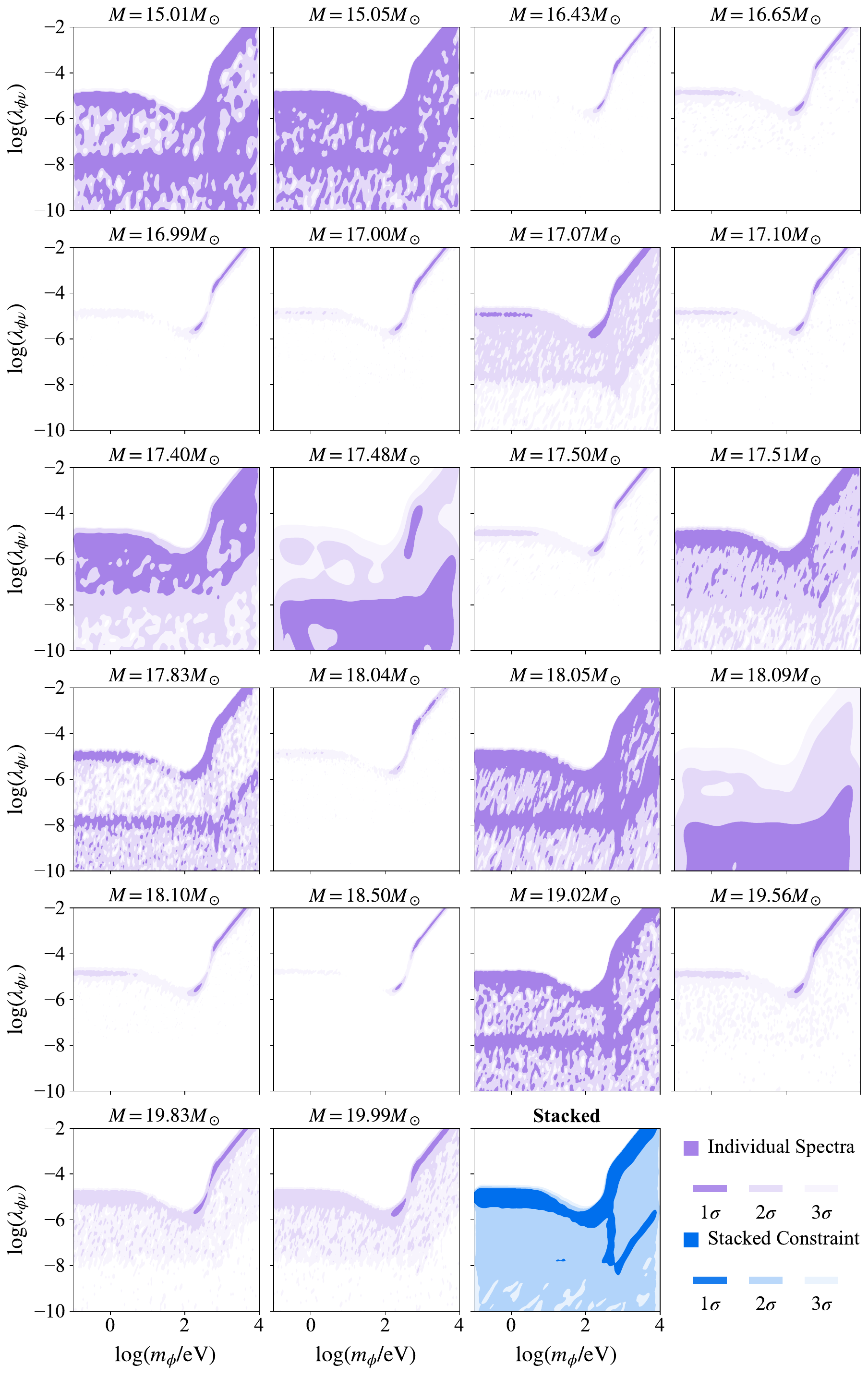}
    \caption{Vartanyan23: 2D contour plots of constraining 2 new physics parameters with 22 independent spectra based on Case III. }
    \label{fig:2d_sim_all_III}
\end{figure*}

For Vartanyan23, we only utilized events before 4s, i.e., 8 events from Kam-II, 6 events from IMB and 3 events from Baksan, to be aligned with the simulation time. We performed independent MCMC procedures based on all 22 spectra of Ref.~\cite{Vartanyan:2023zlb}. Each spectrum represents a pair of effective constraint, as demonstrated in Fig.~\ref{fig:2d_sim_all_I} to \ref{fig:2d_sim_all_III}. These chains are extracted in the same manner as Pagliaroli09, and the constraint for each spectrum differs. For instance, in Case I, spectra simulated with $17.48M_\odot$, $18.09M_\odot$, etc, do not show a strong preference of a specific coupling, and they are consistent with the non-resonant scenario within $1\sigma$ C.L., while most of the spectra prefer a non-zero coupling. However, one cannot rule out the constraint from any mass-based spectrum because we are not certain of the mass of SN1987A. The stacked result is shown in the final contour panels of Fig.~\ref{fig:2d_sim_all_I} to \ref{fig:2d_sim_all_III}.

% \newpage
\begin{figure*}
    \centering
    \includegraphics[width=0.9\linewidth]{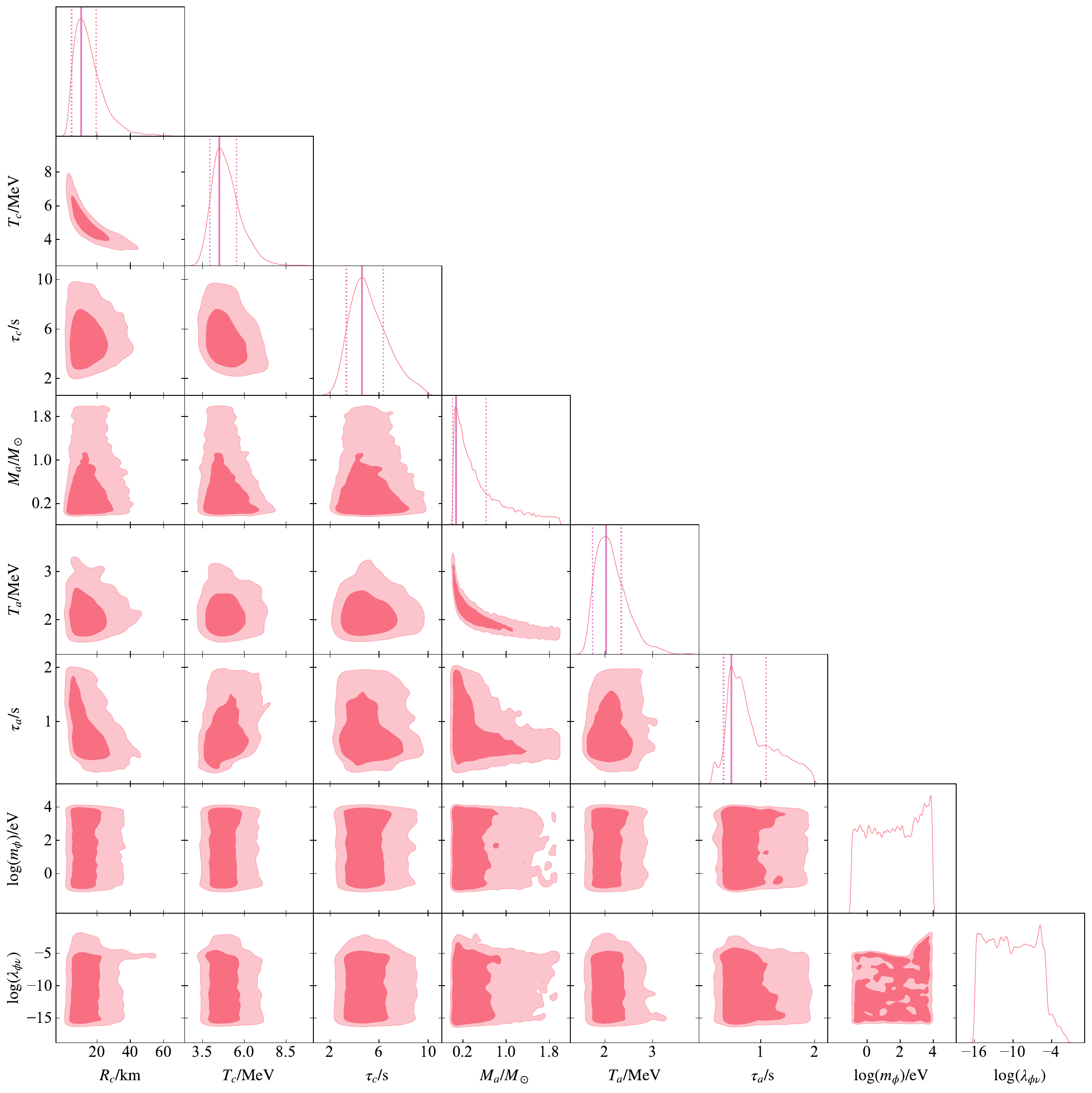}
    \caption{Pagliaroli09: Marginalized posterior distributions and 2D contour plots of constraining 8 parameters simultaneously with current neutrino detections based on Case I. Solid lines in the 1D posteriors denote the best-fit values, while dotted lines indicate the $1\sigma$ C.L. derived by calculating the HPD. Neither best-fit values nor localized $1\sigma$ intervals are applicable to the new physics parameters.}
    \label{fig:8p_I}
\end{figure*}

\begin{figure*}
    \centering
    \includegraphics[width=0.9\linewidth]{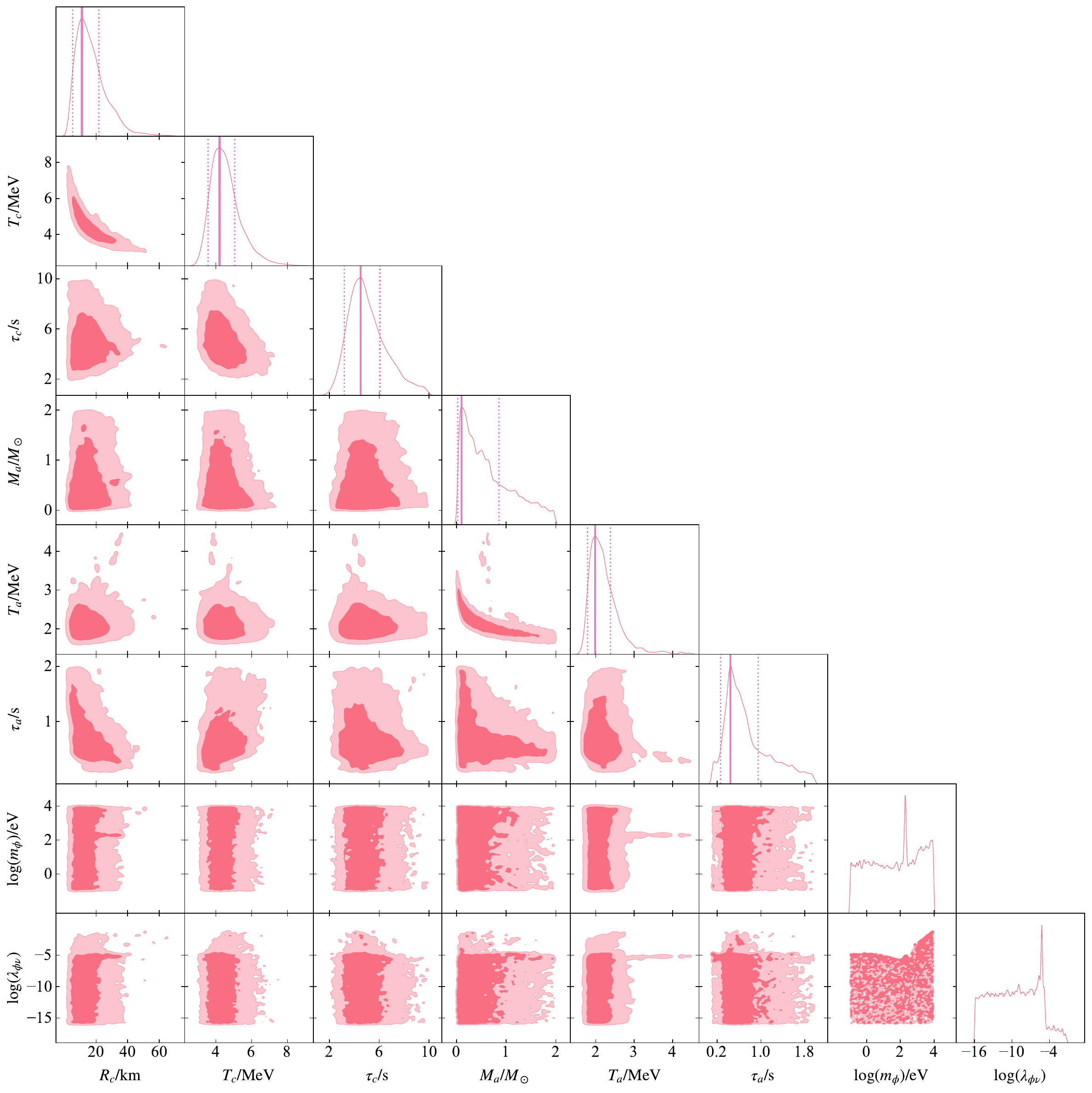}
    \caption{Pagliaroli09: Marginalized posterior distributions and 2D contour plots of constraining 8 parameters simultaneously with current neutrino detections based on Case II. }
\label{fig:8p_II}
\end{figure*}

\begin{figure*}
    \centering
    \includegraphics[width=0.9\linewidth]{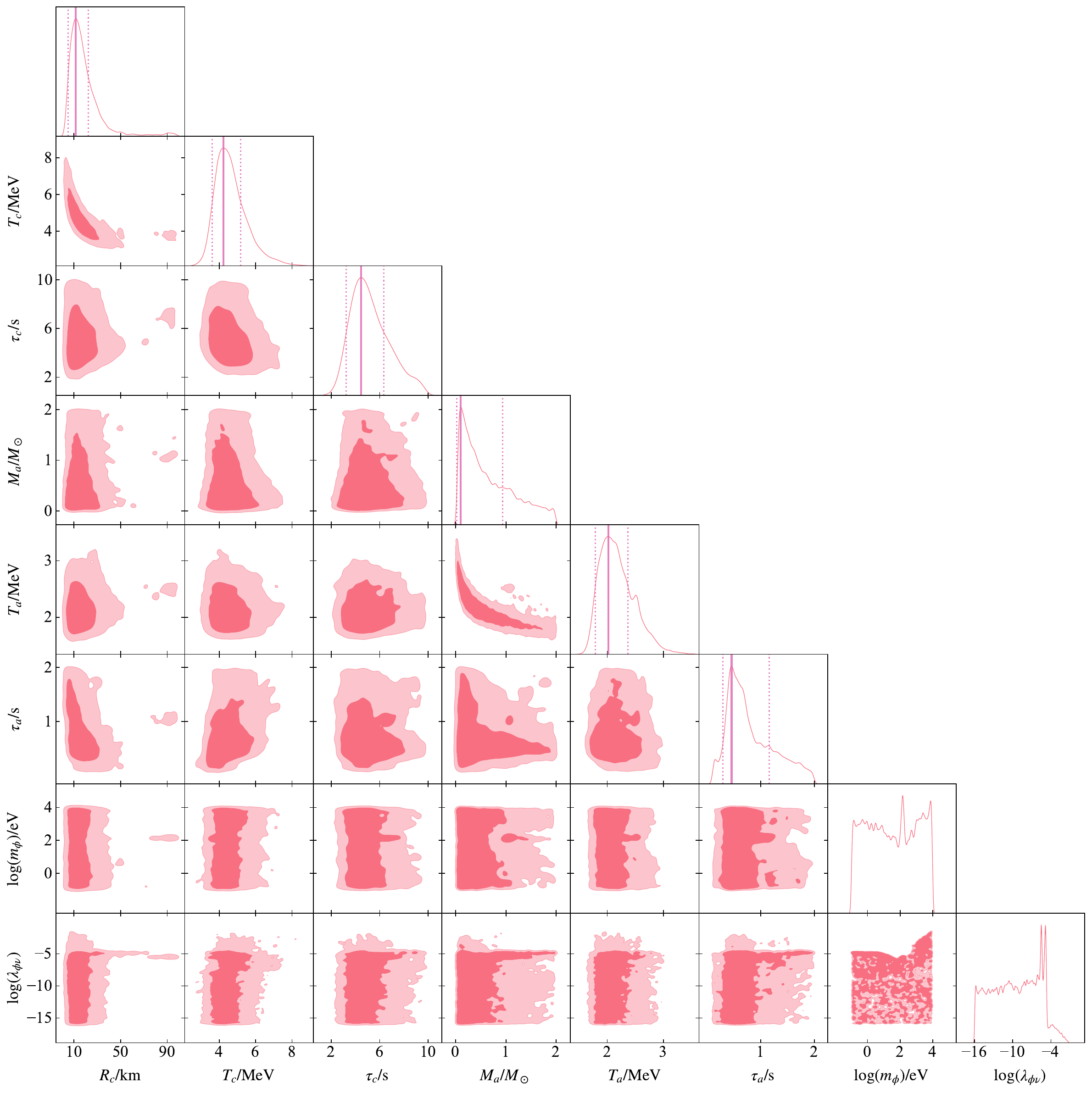}
    \caption{Pagliaroli09: Marginalized posterior distributions and 2D contour plots of constraining 8 parameters simultaneously with current neutrino detections based on Case III. }
\label{fig:8p_III}
\end{figure*}

\bibliography{reference}

\end{document}